\documentclass[a4paper,11pt]{article}

\usepackage{geometry} 

\usepackage{setspace} 

\usepackage{booktabs}
\usepackage{graphicx}
\usepackage{subcaption}
\usepackage{caption}

 \usepackage{caption}
\usepackage{amsmath} 
\usepackage{amssymb}
\usepackage{mathtools}
\usepackage{amsthm}
\usepackage{enumitem}
\usepackage{url} 

\theoremstyle{remark}
\theoremstyle{plain}
\newtheorem{theorem}{Theorem}[section]   
\newtheorem{lemma}[theorem]{Lemma}

\usepackage{booktabs}
\usepackage{multirow}
\usepackage{float}
\usepackage{caption}
\usepackage{xcolor}
\usepackage{pifont}
\usepackage{booktabs}
\usepackage{longtable}
\usepackage{graphicx}
\usepackage{subcaption}
\usepackage{placeins} 
\usepackage{pdflscape} 
\usepackage{amsthm}

\usepackage{booktabs} 

\usepackage[authoryear,round]{natbib} 
\usepackage[colorlinks=true, linkcolor=blue]{hyperref} 
    
\usepackage{comment}

\hypersetup{citecolor=blue,linkcolor=blue,urlcolor=blue} 

\graphicspath{ {./files/}} 

\newcommand{\tablepath}{files} 

\makeatletter
\newcommand{\includetable}[1]{%
  \@ifundefined{tablepath}{%
    \InputIfFileExists{#1}{}{}%
  }{%
    \InputIfFileExists{\tablepath/#1}{}{\InputIfFileExists{#1}{}{}}%
  }%
}
\makeatother

    \makeatother

\begin{document}
\begin{spacing}{1}
\begin{titlepage}
\linespread{1}   
    \begin{center}
        \LARGE
        \textbf{Is There an AI Bubble? Robust Date-Stamping for Periods of Exuberance }
        
        \vspace{0.25cm}
        \Large
        Abir Sarkar\footnote{as4458@cornell.edu, 301 CIS Building,
Ithaca, New York 14853} and Martin T. Wells\footnote{mtw1@cornell.edu, 302 CIS Building,
Ithaca, New York 14853}  
        \vspace{0.25cm}
        
        Department of Statistics and Data Science, Cornell University 
        
        \vspace{0.25cm}
        
      \today
    
        
    \noindent \hrulefill
        
        \begin{abstract}
            \footnotesize{\noindent 
            }
\begin{spacing}{1.2}

The recent surge in valuations among AI-related firms has renewed concerns that markets may be entering a new phase of speculative exuberance, particularly in the technology and semiconductor sectors at the core of the AI investment wave. This paper develops a practical and novel econometric framework for detecting, date-stamping, and drawing inference on the origination and collapse of bubble episodes when prices evolve under persistent, time-varying volatility. Standard bubble-detection methods are typically derived under homoskedasticity or weak forms of heteroskedasticity and may therefore yield misleading inference in more general settings. We extend right-tailed Dickey–Fuller unit root tests to autoregressive models with highly persistent mean and volatility dynamics, delivering a stochastic-volatility-robust ADF (SV-ADF) test that accommodates strongly persistent variance without imposing strict parametric structure. Building on a moderate-deviation asymptotic theory, the SV-ADF yields nuisance-parameter-free procedures with distinct calibration thresholds for testing origination and collapse, producing more stable alarms and fewer transient false positives around volatility spikes. We establish consistency of the date-stamping estimator, show that it remains asymptotically tractable, and demonstrate that it continues to distinguish collapsing bubbles, delivering a practically implementable methodology for monitoring exuberance and retrospectively dating bubble periods. Extensive Monte Carlo simulations document strong power and substantial accuracy over homoskedastic (PWY) procedures when volatility dynamics are pronounced. Our empirical analysis of AI-exposed equities, including the ``Magnificent Seven" and leading semiconductor firms, documents pervasive exuberance with significant cross-sectional heterogeneity in timing, intensity, and duration. The evidence points to particularly strong bubble dynamics for Alphabet and TSMC in the current cycle, while Tesla and Nvidia exhibited especially pronounced explosive episodes during earlier phases of the AI-driven market cycle.

\end{spacing}
        \end{abstract}
 \end{center}
\end{titlepage}


\section{Introduction}\label{section-intro}
\begin{spacing}{1}

The rapid expansion of artificial intelligence has triggered a broad revaluation of large technology and semiconductor firms, as investors have increasingly capitalized expectations of future AI-related cash flows into equity prices. Advances in large language models, cloud computing, data infrastructure, and specialized chip design have placed firms such as Alphabet, Microsoft, Meta, and Nvidia at the center of market attention, while the semiconductor and AI-infrastructure segment, including TSMC and Broadcom more broadly, has become a key transmission channel for AI optimism. This environment is well-suited for bubble dynamics because major innovations raise uncertainty about long‑run payoffs, encourage narrative‑based valuation, and amplify extrapolative beliefs; see \cite{shiller2015irrational}, \cite{pastor2009technological}, \cite{greenwood2019bubbles}.
\begin{figure}[!htpb]
\centering
\includegraphics[width=\linewidth]{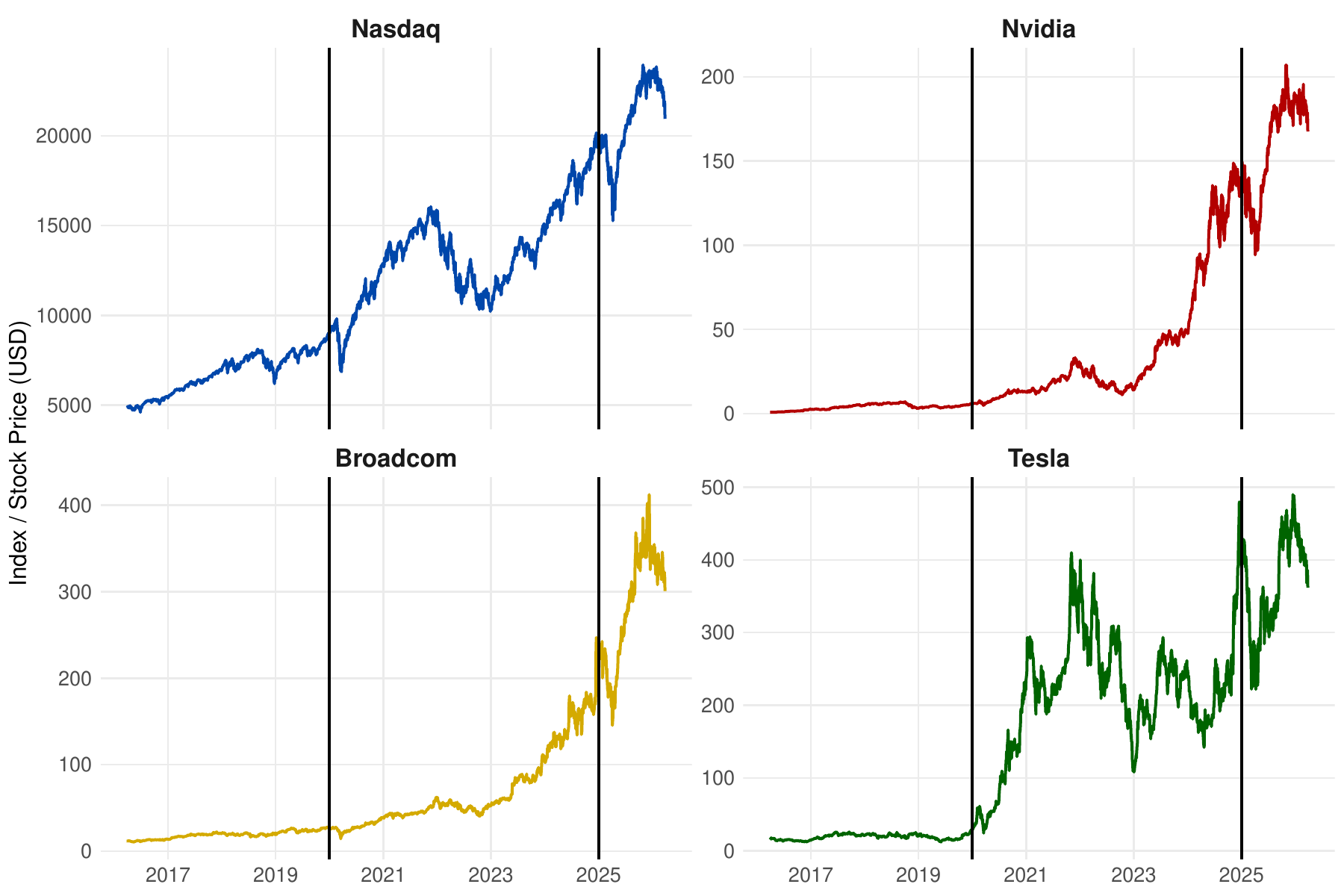}
\captionsetup{skip=0pt}
\caption{\footnotesize The Nasdaq index together with the stock prices of Tesla, Broadcom, and Nvidia over the past ten years. The pronounced run-ups in these stocks during 2020--2025 are suggestive of speculative bubble dynamics. Data source: Yahoo Finance \citep{yahooNVDA}.}
\label{fig:nasdaq-2x2}
\end{figure}

Figure \ref{fig:nasdaq-2x2} compares the stock‑price paths of the Nasdaq index with Nvidia, Broadcom, and Tesla over 2020–2025. While Nasdaq roughly doubled, Nvidia, Broadcom, and Tesla increased by multiple factors. Nasdaq benchmarks broad technology strength, so divergence of individual stocks from the index points to concentrated, firm‑specific dynamics. In the current episode, the price appreciation in AI‑exposed firms has therefore raised a natural question: do these valuations primarily reflect changing fundamentals, or do they also embody speculative exuberance of the kind that has historically preceded sharp market reversals \citep{kindleberger2005manias}.

This concern is not unique to the current AI cycle. Financial history shows that transformative technologies often generate exuberant valuations precisely because their future cash-flow potential is hard to quantify. Railway finance in the 1840s saw a sharp boom followed by a severe reversal as expectations outpaced realizable returns \citep{aliber2015manias, odlyzko2010collective}. Radio and consumer-electronics firms in the 1920s experienced similar valuation surges ahead of the 1929 crash \citep{white1990stock, bruner2019great}. Internet and telecom stocks in the late 1990s likewise rose dramatically before collapsing when growth narratives failed to translate into sustainable cash flows \citep{shiller2000measuring, ofek2003dotcom}. Comparable dynamics appeared outside frontier technology in the U.S. housing boom before the 2007--09 crisis \citep{gorton2008panic, mian2011household, reinhart2009aftermath} and, more recently, in the pandemic-era surge in retail-driven speculative trading \citep{baker2020unprecedented, malz2021gamestop, gormsen2020coronavirus, ramelli2020feverish, barber2022attention}.

Market exuberance and its potential to generate asset-price bubbles have led to a large econometric literature on real-time bubble detection. \citet{phillips2009limit} developed the asymptotic theory for mildly explosive autoregressions, providing a formal foundation to model bubble behavior. \citet{phillips2011explosive} proposed a recursive testing procedure (PWY) for detecting explosive behavior, date-stamping the origination and collapse of exuberance, with an application to the Nasdaq bubble of the 1990s. \citet{phillips2015testing} extended limit theory to real-time dating algorithms (PSY) in environments with multiple bubbles. Since then, a growing literature has applied and extended these methods in both real-time and retrospective settings \citep{harvey2016tests, astill2018real}. Recent work has examined speculative episodes in the “Magnificent Seven”—the group of mega-cap U.S. technology firms comprising Apple, Microsoft, Alphabet, Amazon, Meta, Nvidia, and Tesla, during the AI boom \citep{basele2025speculative}, regional housing markets in Australia, China, Hong Kong, and the United States \citep{shi2016dating, deng2017did, yiu2013detecting, shi2017speculative}, cryptocurrencies \citep{corbet2018datestamping}, common bubbles in large-dimensional financial systems \citep{chen2023common}, and the transmission of cryptocurrency bubbles to systemic risk in global energy companies \citep{ji2022cryptocurrency}.

\begin{figure}[!htb]
\centering
\includegraphics[width=\linewidth]{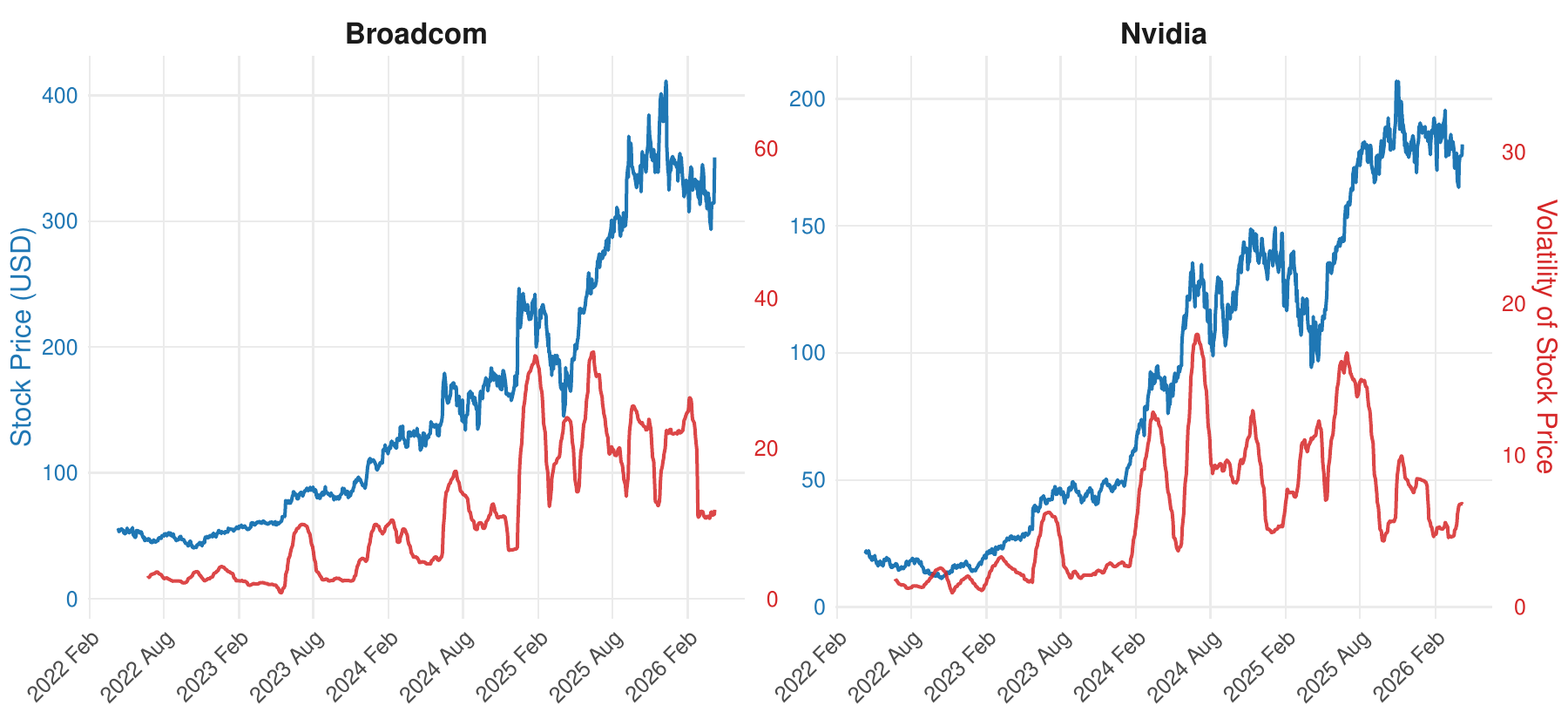}
\captionsetup{skip=0pt}
\caption{\footnotesize Stock price for Broadcom and Nvidia over 2022-2026 exhibit pronounced heteroskedasticity. Volatility is calculated using a rolling window of the previous 8 weeks. The visual evidence supports a specification that allows for persistent variation in scale.}
\label{fig:volatility_4stocks}
\end{figure}

A limitation of the existing bubble-detection literature is its baseline homoskedasticity assumption, which is difficult to reconcile with asset-price data given well-documented volatility clustering, regime shifts, and leverage effects \citep{engle1982autoregressive, engle1986modelling}. Volatility typically rises during periods of exuberance and stress, and high-frequency evidence documents persistent surges in realized volatility around major information events and speculative episodes \citep{greenwood2019bubbles, shephard2010realising, patton2015good}. The stocks most exposed to AI-related narratives exhibit substantial volatility clustering, making the homoskedastic PWY benchmark difficult to sustain. Figure~\ref{fig:volatility_4stocks} provides evidence of these volatility patterns for two major firms prominently exposed to the recent AI era.


The evidence in Figure~\ref{fig:volatility_4stocks} suggests that volatility dynamics must be explicitly modeled rather than treated as a nuisance. This makes the incorporation of time-varying volatility central to credible bubble detection in the current AI era, because otherwise volatility shifts may be mistaken for exuberance and the timing of bubble origination and collapse may be misdated. Recent contributions have begun to incorporate volatility more explicitly into bubble analysis, including de-volatilized recursive tests based on high-frequency measures \citep{boswijk2024testing}, continuous-time no-arbitrage frameworks linking explosive prices to volatility conditions \citep{JarrowProtterShimbo2010}, option-implied approaches for measuring bubble magnitudes under stochastic volatility \citep{jarrow2021inferring, fusari2025testing}, quadratic-variation methods for identifying explosive behavior \citep{jarrow2024study}, and procedures for dating cryptocurrency bubbles and evaluating gains relative to GARCH benchmarks \citep{choi2020testing, montanino2025bubble}. Given the pronounced run-ups and intermittent corrections in AI-exposed equities, together with elevated volatility around earnings announcements, product releases, and policy news, allowing for stochastic, time-varying volatility is essential for credible bubble detection and date-stamping in the current AI era.

\subsection{Our Contributions}
Motivated by the pronounced run-ups in AI-exposed technology and semiconductor stocks, and by the evident time variation in their volatility, this paper makes three major contributions to the literature on asset-price bubble detection and inference.

\begin{figure}[!htpb]
\centering
\includegraphics[width=\linewidth]{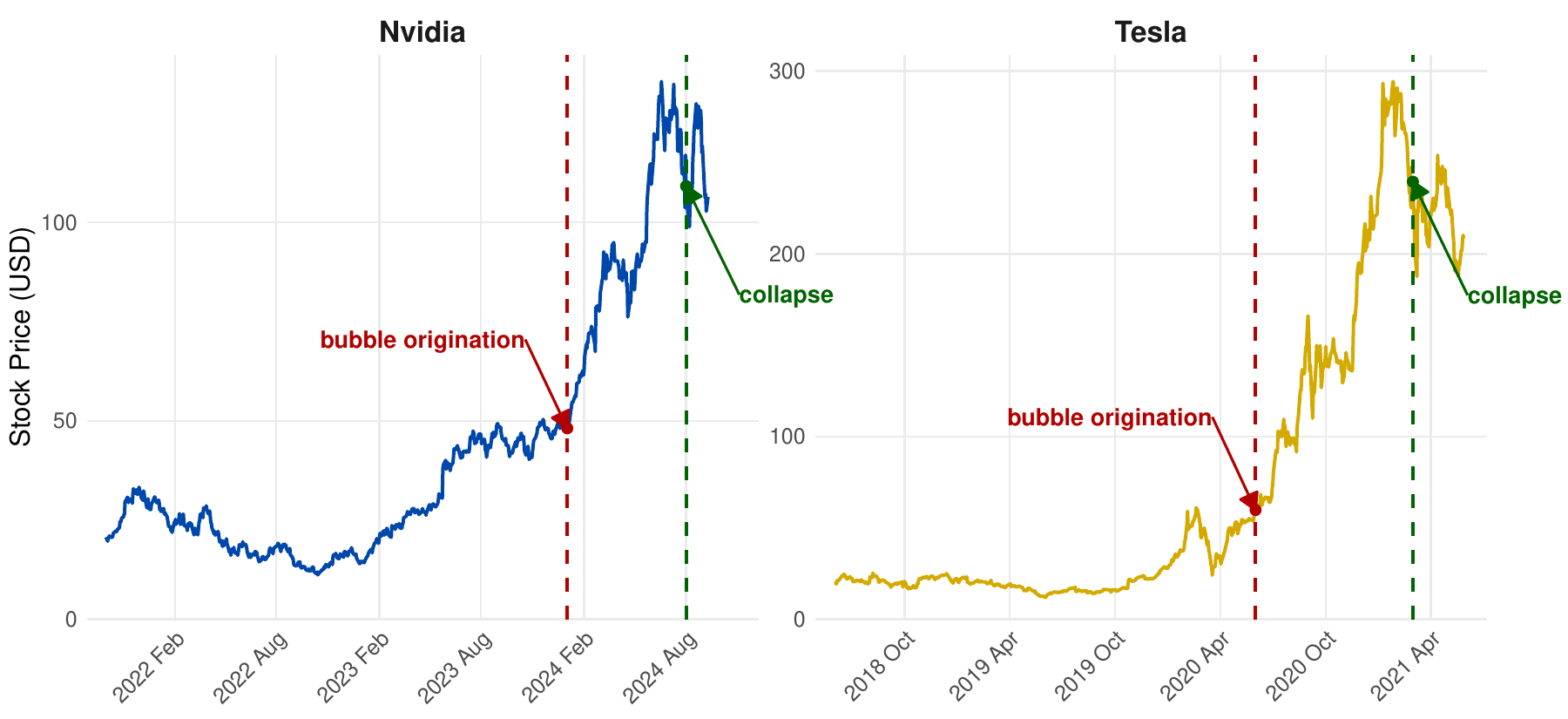}
\captionsetup{skip=0pt}
\caption{\footnotesize Proposed SV-ADF procedure identified speculative explosive bubble behaviour in individual stocks. The dotted red vertical lines mark bubble origination, the dotted green lines signify collapse. Speculative intervals: Nvidia (Jan' 2024 -- Aug' 2024), Tesla (June 2020--March 2021).}
\label{fig:bubble_intro}
\end{figure}

First, we develop a novel volatility‑robust stochastic ADF test and its recursive sup‑type implementation for retrospective bubble date-stamping, as shown in Figure~\ref{fig:bubble_intro}. The procedures accommodate double near‑unit behavior, mild drift toward explosiveness in the mean with highly persistent  volatility  \cite{sarkar2025double}, without requiring a tightly specified volatility model. We also provide confidence intervals for the autoregressive coefficient and for explosive growth rates that remain valid under persistent volatility.

\begin{figure}[!htpb]
\centering
\includegraphics[width=\linewidth]{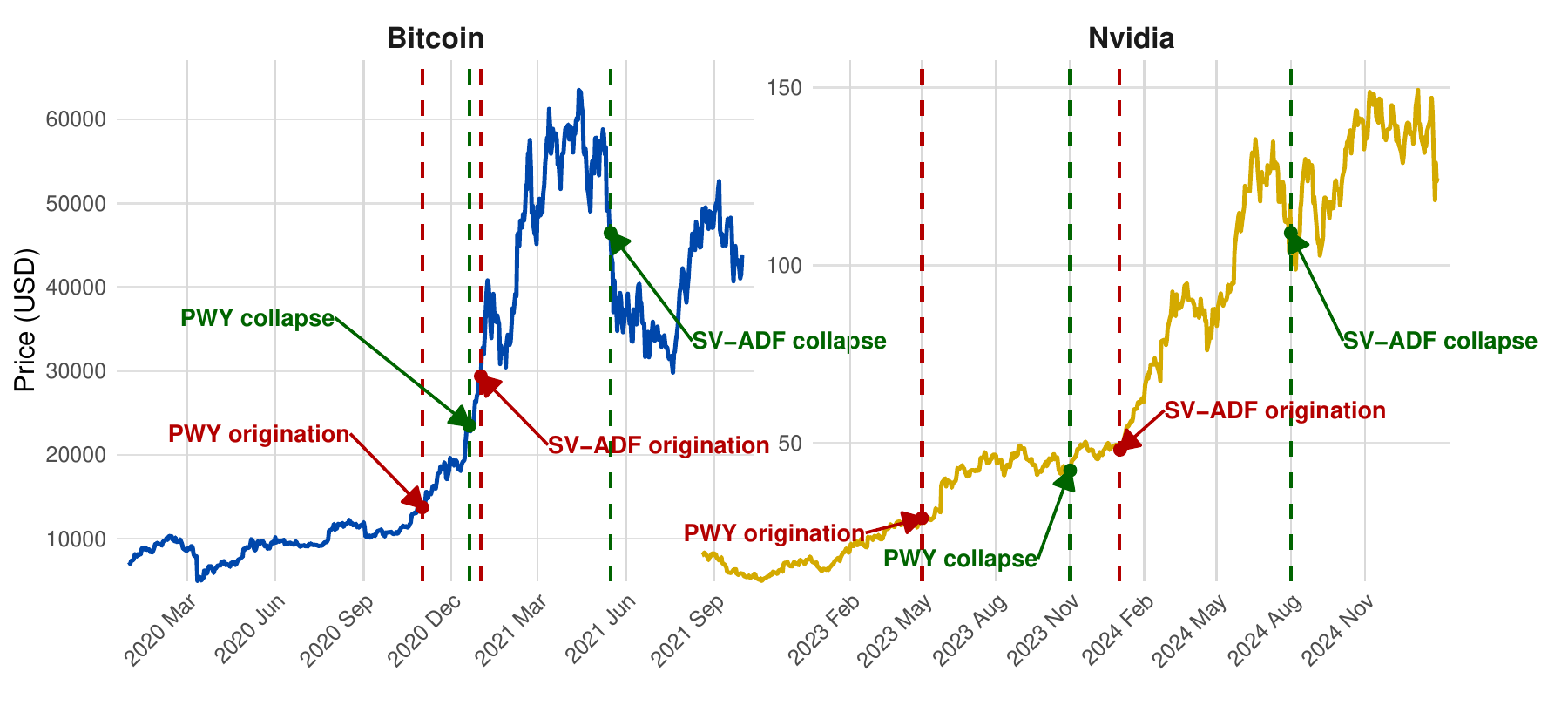}
\captionsetup{skip=0pt}
\caption{\footnotesize Misdating of Bitcoin and Nvidia bubble episodes under the PWY approach. This discrepancy likely reflects the reliance of homoskedastic thresholds in an environment where volatility is strongly time-varying, as is typical in cryptocurrency markets. SV-ADF procedure delivers date-stamps that are more economically plausible, aligned with major news developments.}
\label{fig:bubble_misdating}
\end{figure}

Second, we show that bubble origination and collapse should not share the same decision threshold, contrary to conventional PWY practice. A unit-root-based threshold suits origination because the pre-bubble regime is non-explosive. For collapse, the process has already passed through an explosive phase, so reusing the same cutoff is not theoretically justified. Our framework addresses this asymmetry by calibrating the two thresholds separately. This also resolves a deeper structural limitation of PWY. Collapse is modeled through an explicit re-initialization mechanism that forces the post-bubble process back to an $O_p(1)$ neighborhood of the pre-bubble level. This is economically restrictive since, under mild explosiveness, the terminal bubble level typically grows exponentially (see the \hyperref[sec:pwy_contra]{Appendix A.6}). By contrast, our approach imposes no such constraint and lets collapse be inferred from the data's recursive behavior. The resulting asymptotic theory is more general than PWY and more credible for the large run-ups observed in AI-exposed equities. In simulations with persistent volatility, it delivers higher power and more accurate date-stamping while avoiding over-classification when volatility rises without genuine explosiveness. Figure~\ref{fig:bubble_misdating} illustrates this difference for Bitcoin and Nvidia, and Table~\ref{tab:bubble_decisions_two_periods} shows that in recent AI data, PWY often classifies too many equities as exuberant, whereas SV-ADF is more selective and economically coherent.

Finally, we provide an inferential theory to consistently date-stamp the bubble start and end periods. In AI‑exposed equities, SV‑ADF date‑stamps align with earnings‑driven regime shifts, and produce more stable origination signals with fewer transient false alarms. Figure~\ref{fig:bubble_intro} provides the visual evidence of exuberance in Nvidia and Tesla stock prices, and SV-ADF procedure date-stamps the start and end time-periods. For the 1990s Nasdaq, our procedure dates bubble origination to April 1995, earlier than the June 1995 date implied by PWY, despite employing conservative thresholds, and identifies collapse in September 2000. For Bitcoin and Ethereum, standard PWY tests misdate the bubble period, as documented by \citet{boswijk2024testing}, whereas the SV-ADF identifies bubble windows that closely match the results from de-volatilized recursive tests. Robustness holds across sampling frequencies and volatility measures.


The remainder of the paper is organized as follows. Section~\ref{sec:notations}  lists all mathematical notations used throughout this paper. Section~\ref{section-methods}  motivates the empirical setting by documenting the concentration of AI-exposed firms, identifying the most relevant time window for bubble analysis. Section~\ref{sec:section3} develops the SV–ADF test under mild conditions on volatility process. Section~\ref{sec:section4} presents asymptotic theory and confidence intervals for the autoregressive coefficient and explosive growth rates and establishes asymptotic results for the consistency of date‑stamping bubble origination and collapse. Section~\ref{sec:simulation} reports simulation and real data evidences, the findings indicate pronounced exuberance with meaningful intensity in some assets for significant duration, and the estimated windows align closely with major market and corporate announcements, supporting the practical relevance of the methodology. Section~\ref{sec:conclusion} concludes with a summary of contributions and directions for future research. Proofs of the main results and supplementary analyses are collected in the \hyperref[sec:appendix]{Appendix}.

\subsection{Mathematical Notations}\label{sec:notations}
All random quantities are defined on a common probability space
\((\Omega,\mathcal F,\mathbb P)\). We write \(\mathbb P\) for probability and
\(\mathbb E\) for expectation. Throughout, the filtration $
\mathcal F_t
:=
\sigma\!\bigl(X_0,\varepsilon_1,\ldots,\varepsilon_t,\eta_1,\ldots,\eta_t\bigr)
$
denotes the information available up to time \(t\). For deterministic sequences, \(a_n=o(b_n)\) means \(a_n/b_n\to 0\),
\(a_n=O(b_n)\) means \(|a_n/b_n|\) is uniformly bounded,
\(a_n\sim b_n\) means \(a_n/b_n\to 1\), and \(a_n\asymp b_n\) means that
\(a_n/b_n\) is bounded away from both \(0\) and \(\infty\).
For random sequences, \(X_n=o_p(1)\) means \(X_n\xrightarrow{p}0\),
\(X_n=O_p(1)\) means \(\{X_n\}\) is bounded in probability,
\(X_n\xrightarrow{p}X\) denotes convergence in probability, and 
\(X_n\Rightarrow X\) denotes convergence in distribution. We use \(\mathbf 1\{\cdot\}\) for the indicator function,
\(\lfloor\cdot\rfloor\) for the integer part, and \(\log(\cdot)\) for the natural
logarithm. The notation \(\mathcal N(\mu,\sigma^2)\) denotes a normal
distribution with mean \(\mu\) and variance \(\sigma^2\), and
\(\mathcal C\) denotes the standard Cauchy law.

\end{spacing}

\section{Price Dynamics of AI-Exposed Tech and Semiconductors}\label{section-methods}
\begin{spacing}{1}
The current AI cycle is distinguished not only by rapid technological progress, but also by the extraordinary concentration of market value in a small group of firms perceived to be central to the AI economy. This concentration is especially visible among the so‑called “Magnificent Seven” firms, analyzed in \cite{basele2025speculative}. As of March 2026, the market capitalizations of these firms are shown in Table~\ref{tab:ai_caps}. A similar pattern appears within the semiconductor and AI‑linked infrastructure segment, where Nvidia, TSMC, Broadcom, ASML, Micron, and Palantir account for a substantial share of aggregate valuations. These magnitudes underscore how strongly the current AI episode is tied to the valuation of a narrow but economically dominant set of firms.

\begin{table}[!htpb]
\centering
\begin{tabular}{l r @{\hspace{2.5em}} l r}
\toprule
\multicolumn{2}{c}{Large Technology} & \multicolumn{2}{c}{Semiconductors and AI‑linked infrastructure} \\
\cmidrule(r){1-2}\cmidrule(l){3-4}
Company & Market cap & Company & Market cap \\
\midrule
Apple      & \$3.83T & Nvidia    & \$4.49T \\
Alphabet   & \$3.71T & TSMC      & \$1.80T \\
Microsoft  & \$3.02T & Broadcom  & \$1.62T \\
Amazon     & \$2.30T & ASML      & \$0.54T  \\
Meta       & \$1.65T & Micron    & \$0.45T  \\
Tesla      & \$1.50T & Palantir  & \$0.36T  \\
\bottomrule
\end{tabular}
\caption{Market capitalizations as of March 2026. Figures are approximate and reported in U.S. dollars. Source: Yahoo Finance \citep{yahooNVDA}.}
\label{tab:ai_caps}
\end{table}

Potential bubble dynamics in these firms are of first-order importance because their large index weights and broad portfolio exposure imply that any detachment from fundamentals could distort aggregate valuations and heighten vulnerability to abrupt corrections. Over 2020--2025, the Nasdaq index roughly doubled, yet several AI-exposed stocks appreciated by far larger multiples. In our sample, Nvidia rose from roughly \$18 in January 2023 to \$147 in November 2024, TSMC from about \$64 in October 2022 to over \$300 in December 2025, and Broadcom from about \$47 in November 2022 to \$325 by October 2025. Such price paths naturally invite comparison with earlier episodes of market exuberance, including the late-1990s Nasdaq boom \citep{phillips2011explosive} or subprime crisis documented by \cite{phillips2011dating}. These considerations motivate our focus on the recent AI era as the most relevant setting for testing whether recent AI-driven price run-ups reflect episodes of explosive bubble behavior. 

\subsection{Window and Data Frequency Selection}

\begin{figure}[!htpb]
\centering
\includegraphics[width=\linewidth]{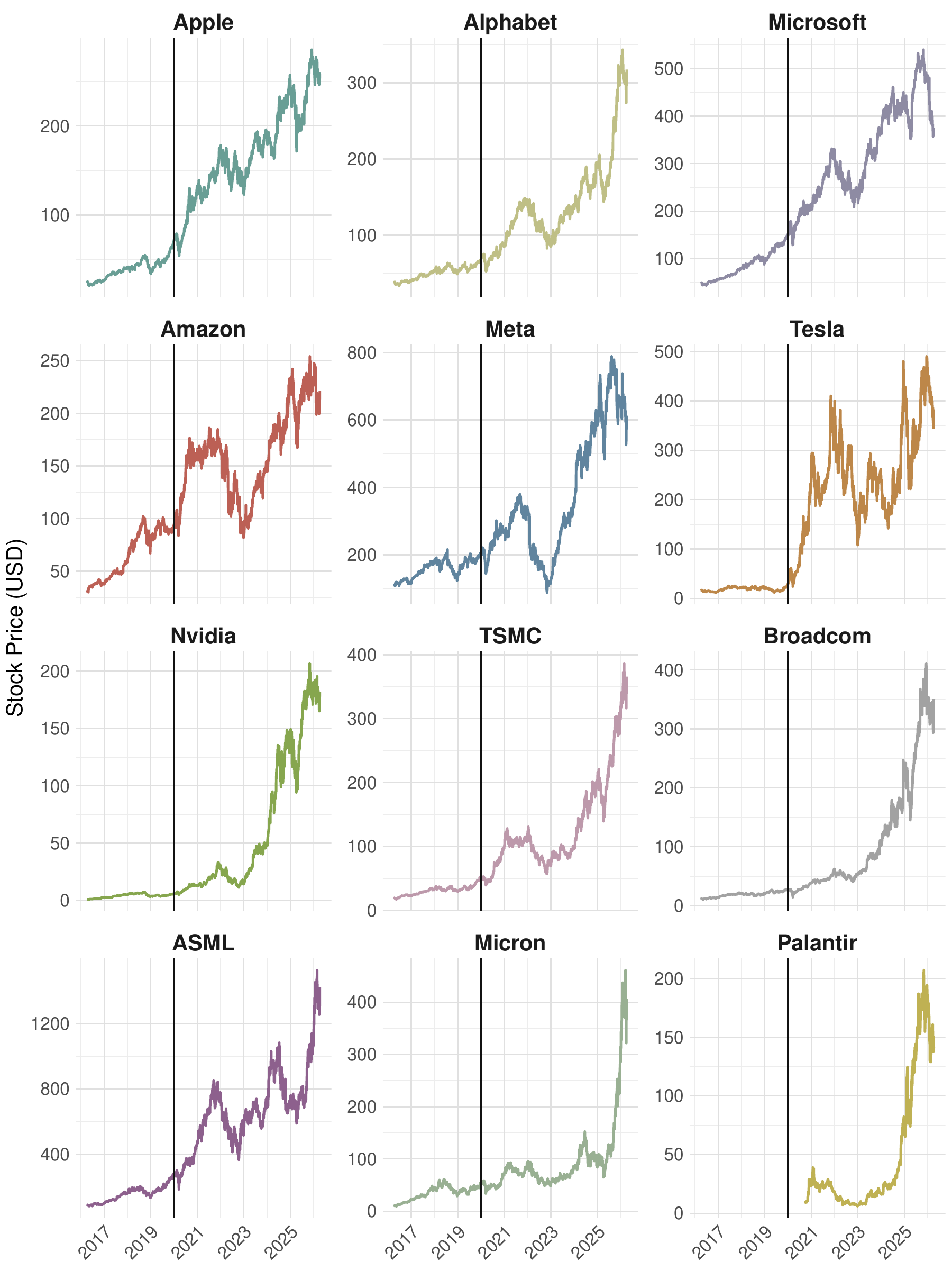}   
\captionsetup{skip=0pt}
\caption{\footnotesize  Daily adjusted close prices of the Magnificent 7 and the six largest semiconductor/AI-infrastructure firms by market capitalization over the past 10 years. Panels are arranged in descending market-cap order, with the first six showing tech firms and the next six for the rest. The vertical line at January 2020 marks the onset of the surge emphasized in our analysis.  Source: Yahoo Finance \citep{yahooNVDA}.}
\label{fig:12stocks} 
\end{figure}

We begin with daily adjusted prices over the past ten years for the twelve firms listed in Table~\ref{tab:ai_caps}. The ten-year horizon provides a broad benchmark against which the recent AI episode can be evaluated, while the daily frequency is chosen because our bubble-dating procedure is recursive and event-sensitive, so finer granularity preserves information relevant for the SV-ADF tests. Figure~\ref{fig:12stocks} displays these price paths over the full sample. Most pre-2020 observations correspond to a comparatively benign regime in which AI-related cash-flow narratives and computation bottlenecks were still nascent, so long stretches of near-unit-root behavior contribute relatively little to the identification of recent explosive episodes. By contrast, the 2020--2025 window encompasses the COVID shock and rebound, the retail-attention episodes documented in the recent literature, and the late-2022 shift associated with the public rollout of generative-AI applications and the ensuing surge in GPU demand \citep[see][]{baker2020unprecedented, malz2021gamestop, barber2022attention}. For this reason, the empirical analysis focuses primarily on the post-2020 period, where AI-exposed equities display the most plausible candidate windows for exuberance.

\subsection{Economic Motivation for Firm-Level Bubble Analysis}

The composition of the sample is economically motivated, since AI optimism is transmitted through distinct valuation channels across platform firms and infrastructure bottlenecks. The large technology firms in our sample are natural candidates for bubble analysis in the current AI era because their valuations increasingly reflect expectations about long-run AI monetization rather than current earnings alone. For Alphabet, Microsoft, Amazon, Meta, and Apple, these expectations operate through AI integration into search, cloud computing, advertising, enterprise software, and consumer ecosystems. Tesla, while distinct from the platform firms, remains equally relevant because its valuation has repeatedly been shaped by broader narratives of technological transformation, including autonomy, robotics, and AI-enabled mobility.

The semiconductor segment is even more central because it contains the key physical and technological bottlenecks through which the AI boom is transmitted into market valuations. Firms in this segment span the computational, fabrication, networking, memory, and software layers of the AI value chain, making them especially informative for bubble analysis. Nvidia and Broadcom are central to AI computation and infrastructure, while TSMC and ASML occupy critical bottleneck positions in advanced chip fabrication. Micron has become increasingly important through high-bandwidth memory, and Palantir provides a software-side proxy for the commercialization of AI in enterprise and government settings. Taken together, these firms offer a natural laboratory for assessing whether markets are pricing durable fundamentals, speculative narratives, or some combination of the two.

\newenvironment{assumptions}{%
\par\noindent\textbf{Assumptions.}%
\begin{enumerate}%
\renewcommand\labelenumi{A\arabic{enumi}.}%
\renewcommand\theenumi{A\arabic{enumi}.}
}{%
\end{enumerate}%
}

\end{spacing}

\newpage

\section{Explosive Bubble under Stochastic Volatility}\label{sec:section3}
\begin{spacing}{1}
\cite{phillips2009limit} proposed recursive regression methods for detection of  bubble behavior, thereby providing a formal econometric foundation for testing asset price exuberance. Very recently,  \cite{sarkar2025double} established a unified moderate-deviation asymptotic theory for autoregressive models with jointly persistent mean and volatility dynamics. The central idea in this literature is to employ right-sided unit root tests to detect mildly explosive behavior, in contrast to the conventional left-sided unit root tests designed to detect stationarity, using either the estimated autoregressive coefficient or the associated $t$-statistic. We next develop a recursive ADF procedure within this stochastic-volatility framework for the purpose of testing and date-stamping explosive behavior.

\newtheorem{assumption}{Assumption} 

\noindent For each asset price series $\{x_t\}_{t=1}^n$, we estimate by least squares the augmented autoregression
\begin{equation}
x_t \;=\; \mu \;+\; \delta\,x_{t-1} \;+\; \sum_{j=1}^{L}\phi_j \,\Delta x_{t-j} \;+\; u_t,
\qquad
u_t \;=\; \sigma_t \varepsilon_t,\;\; \varepsilon_t \stackrel{\text{i.i.d.}}{\sim}\mathcal{N}(0,1),
\label{eq:ar-aug}
\end{equation}
where $L$ is determined by a sequential significance test on the coefficients $\{\widehat \phi_j\}_{j=1}^L$. We do not impose a parametric model on $\{\sigma_t\}$; instead, our asymptotic results for the SV–ADF statistics require only the mild regularity conditions stated below. Define the average innovation variance up to time $k$ by
\begin{equation}
\newcommand{\E}{\ensuremath{\mathbb{E}}}
m_{n,k} \;:=\; \frac{1}{k}\sum_{t=1}^{k} \E\,\!\big(\sigma_t^{2}\big),
\label{eq:mnk}
\end{equation}
which serves as the effective volatility scale when $\sigma_t^2$ is time‑varying.

\begin{assumption}\label{ass:LLNscale}
For \(r\in[r_0,1]\), let \(k=\lfloor nr\rfloor\). Then, for every \(r_0\in(0,1)\), as $ n\to \infty$,
\[
\sup_{r\in[r_0,1]}
\left|
\frac{1}{k\,m_{n,k}}
\sum_{t=1}^{k}\sigma_t^2
-1
\right|
\xrightarrow{p}0.
\]
\end{assumption}

\begin{assumption}
\label{ass:smooth}
There exists some $\zeta \in (0,1)$ such that \[\sup_{t>n^\zeta}\left|\frac{m_{n,t}}{m_{n,n}}-1\right| = o\!\left(n^{-1/2}\right).\]
\end{assumption}

Note that, these assumptions are considerably weaker than homoskedasticity. They permit highly persistent, nearly nonstationary stochastic volatility (e.g., $\log\sigma_t^2=\phi_n\log\sigma_{t-1}^2+\eta_t$ with $\phi_n\to 1$ at an iterated logarithmic rate; see \cite{sarkar2025double}). The canonical AR(1) log-volatility specifications \citep{taylor2008modelling, harvey2016tests}, GARCH‑type processes with $\alpha+\beta$ close to but less than one \citep{bollerslev1994arch,baillie1996fractionally}, slowly varying deterministic variance \citep{dahlhaus1997fitting, cavaliere2008bootstrap} can also be allowed under the proposed model specifications. 

Assumption \ref{ass:LLNscale} ensures that the predictable quadratic variation of the martingale array $\{u_t\}$, scaled suitably, converges to a non-trivial limit. This delivers a functional central limit theorem for the partial‑sum process $W_n(\cdot)$ and underpins the Brownian functional limits in the SV-ADF statistic. 
 The uniform $o(n^{-1/2})$ control in Assumption~\ref{ass:smooth} for $t>n^{\zeta}$ ensures that, beyond an initial fraction of the sample, the expected scale varies slowly enough that residual drift does not affect the asymptotic behavior of the recursive SV-ADF statistic. In this sense, the condition is substantially weaker than homoskedasticity while still strong enough to support volatility-robust limits. More broadly, the recursive SV-ADF literature has been developed without imposing the restrictive $O_p(1)$ assumption as done in homoskedastic PWY approach, so extending it to settings with highly persistent and systematically evolving stochastic volatility is a central novel feature of our approach.

We adopt a forward recursive regression framework and re-estimate model \eqref{eq:ar-aug} on expanding subsamples. The null is $H_0:\,\delta=1$ against the right‑tailed alternative $H_1:\,\delta>1$. If the first regression uses $\tau_0=\lfloor n r_0\rfloor$ observations for some $r_0\in(0,1)$, subsequent regressions augment this initial sample by one observation at each step, yielding $\tau=\lfloor n r\rfloor$ for $r\in[r_0,1]$. Let
\begin{equation}
\widetilde X_{j-1}
\;:=\;
X_{j-1}
-
\frac{1}{\tau}\sum_{k=1}^{\tau}X_{k-1},
\qquad j=1,\dots,\tau,
\label{eq:Xtilde-def}
\end{equation}
and define the forward recursive SV-ADF statistics
\begin{equation}\label{eq:normalized_stats}
\text{SV-ADF}_r^{\delta} \;:=\; \tau\bigl(\widehat\delta(\tau)-1\bigr), 
\qquad
\text{SV-ADF}_r^{t} \;:=\; 
\left( \frac{\sum_{j=1}^{\tau}\widetilde X_{j-1}^{\,2}}{\widehat\sigma_{\tau}^{2}} \right)^{1/2}
\bigl(\widehat\delta(\tau)-1\bigr),
\end{equation}
where $\widehat\delta(\tau)$ is the least squares estimator of $\delta$ based on the first $\tau$ observations and $\widehat\sigma_{\tau}^{2}$ is a consistent estimator of the average innovation variance over the same window. 

To obtain a feasible $t$‑type statistic, the unknown average volatility scale $m_{n,\tau}$ must be replaced by a sample‑based estimator. The following condition ensures that the recursive variance estimator $\widehat\sigma_\tau^2$ consistently recovers this scale uniformly over the recursion range.

\begin{assumption}\label{ass:sigmahat}
For $r\in[r_0,1]$
$$\frac{\widehat\sigma_{\lfloor n r\rfloor}^{2}}{m_{n,\lfloor n r\rfloor}}\ \xrightarrow{p}\ 1.$$
\end{assumption}
\noindent The next theorem shows that the forward recursive SV-ADF statistics retain tractable Brownian functional limits under the unit-root null even when volatility is stochastic and highly persistent.
\begin{theorem}\label{thm:SVADF-null}
Suppose $H_0:\,\delta=1$ in \eqref{eq:ar-aug}, $X_0=o_p\,\!\big(\sqrt{n\,m_{n,n}}\big)$, and Assumptions \ref{ass:LLNscale}–\ref{ass:sigmahat} hold. Let $W$ be standard Brownian motion on $[0,1]$, define the demeaned bridge on $[0,r]$ by
\[
\widetilde W_r(u)\;:=\;W(u)\;-\;\frac{1}{r}\int_{0}^{r}W(s)\,ds,\qquad u\in[0,r],\ \ r\in[r_0,1].
\]
Then, as $n\to\infty$ with $\tau=\lfloor n r\rfloor$ and $r\in[r_0,1]$,
\begin{align}
\text{SV-ADF}_r^{\delta}
&\ \Rightarrow\ 
r \,\frac{\displaystyle \int_{0}^{r} \widetilde W_r(u)\, dW(u)}
     {\displaystyle \int_{0}^{r} \widetilde W_r(u)^{2}\, du}, 
\label{eq:DFdelta-limit-thm}\\[0.4em]
\text{SV-ADF}^{\,t}_r 
&\ \Rightarrow\ 
\frac{\displaystyle \int_{0}^{r} \widetilde W_r(u)\, dW(u)}
     {\displaystyle \left(\int_{0}^{r} \widetilde W_r(u)^{2}\, du\right)^{1/2}}.
\label{eq:DFt-limit-thm}
\end{align}
\end{theorem}

Theorem~\ref{thm:SVADF-null} provides the asymptotic foundation for recursive bubble testing under a wide class of heteroskedastic volatility specifications. By establishing nuisance-parameter-free volatility-robust limits for the forward recursive SV-ADF statistics, it delivers a valid basis for critical-value construction and sequential testing in environments with time-varying volatility.

\begin{figure}[!htb]
\centering
\includegraphics[width=\linewidth]{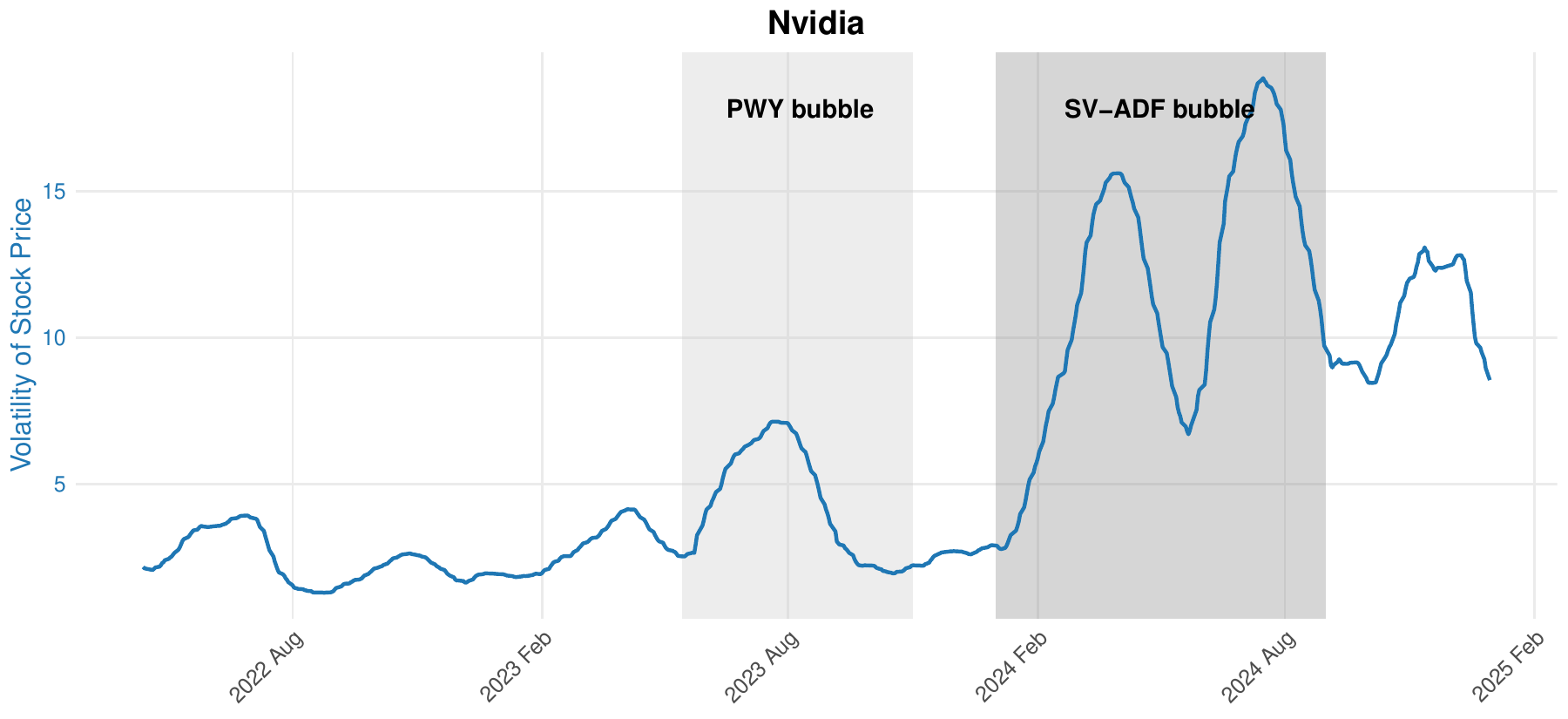}
\captionsetup{skip=0pt}
\caption{\footnotesize Stock-price volatility for Nvidia over 2022--2025 exhibits pronounced peaks and troughs. PWY signals a spurious bubble window in one such high-volatility episode (June--November 2023). SV-ADF procedure identifies the most persistent and quantitatively pronounced episode of explosiveness (January--August 2024), rather than short-lived price spikes induced by volatility fluctuations.}
\label{fig:volatility_misdating}
\end{figure}

An immediate application of Theorem~\ref{thm:SVADF-null} is for the calibration of recursive SV-ADF tests used in bubble dating. Although after normalization by the average volatility scale, the limiting functionals coincide with those obtained under homoskedasticity \citep{phillips2009limit}, the finite-sample critical values need not. In particular, thresholds calibrated under homoskedasticity can be misaligned when volatility is time-varying, which may lead to spurious bubble origination signals in high-volatility periods. By contrast, Theorem~\ref{thm:SVADF-null} provides a basis for volatility-consistent critical values and avoids reliance on a homoskedastic bootstrap approach \citep{phillips2020real}, thereby preserving size along the recursion. Figure~\ref{fig:volatility_misdating} provides evidence of spurious bubble signals by PWY in Nvidia stock price during volatility fluctuations.  We develop the bubble-dating algorithm and its implications for origination and collapse inference in the next section.

\end{spacing}


\section{Estimation and Inference of Bubble}\label{sec:section4}
\begin{spacing}{1}

\end{spacing}
We adopt the standard representation in which the presence of a bubble is captured by mildly explosive dynamics in the autoregressive root. Following \citet{phillips2007limit, magdalinos2009econometric} and \citet{sarkar2025double}, consider the AR(1) model
\begin{equation}\label{eq:main_model_used}
y_t \;=\; \delta_n\,y_{t-1} \;+\; \sigma_t \varepsilon_t,
\qquad
\varepsilon_t \stackrel{\mathrm{i.i.d.}}{\sim} (0,1),
\end{equation}
with mildly explosive coefficient
\begin{equation}
\delta_n \;=\; 1 \;+\; \frac{c}{n^{\alpha}},
\qquad c>0,\;\; \alpha\in(0,1).
\label{eq:me-root}
\end{equation}
Under standard regularity conditions, least-squares inference in this setting exhibits self-normalized Cauchy limits. Writing $\widehat\delta_n$ for the OLS estimator of $\delta_n$, one has
\[
\left(\frac{n^{\alpha}\,\delta_n^{\,n}}{2c}\right)\bigl(\widehat{\delta}_n-\delta_n\bigr)\;\Rightarrow\;\mathcal{C},
\]
where $\mathcal{C}$ denotes a standard Cauchy random variable. Formally, inversion of this limit suggests confidence regions for $\delta_n$ of the form
\[
\widehat{\delta}_n
\;\pm\;
\left(\frac{2c}{n^\alpha\,\delta_n^{\,n}}\right)c_\beta^{\mathcal C},
\]
where $c_\beta^{\mathcal C}$ denotes the two-sided quantile corresponding to coverage level $\beta$ under the standard Cauchy law. Since $\delta_n$ is unknown, this expression is not directly feasible. However, the parametric structure in \eqref{eq:me-root} permits one to leverage maximum likelihood estimation type arguments to construct
 asymptotically valid feasible intervals based only on $\widehat\delta_n$. Combining the arguments in \citet{phillips2023estimation} and \citet{sarkar2025double}, we obtain the following confidence intervals.

\noindent If $\widehat\delta_n$ denotes the least-squares estimator of $\delta_n$ based on $\{y_t\}_{t=1}^n$, define
\begin{equation}
\widehat \gamma_n \;:=\; -\,\frac{\log|\widehat\delta_n-1|}{\log n}.
\label{eq:theta-gamma-hat}
\end{equation}
The quantity $\widehat\gamma_n$ indexes the estimated local deviation of the autoregressive root from unity and therefore summarizes the implied growth rate of local explosiveness. From the perspective of bubble estimation, inference on $\delta_n$ and $\gamma_n$ serves two complementary roles. The confidence interval for $\delta_n$ determines whether the autoregressive root is statistically separated from unity and therefore whether the price process exhibits explosive behavior. The confidence interval for $\gamma_n$ quantifies the intensity of explosiveness and the uncertainty surrounding it.

\noindent If $\widehat\delta_n<1$, an asymptotic $100\beta\%$ confidence interval for $\delta_n$ is
\begin{equation}
\widehat\delta_n \;\pm\; c_{\beta}\,\frac{2}{n^{(1+\widehat\gamma_n)/2}},
\qquad
c_{\beta} \;=\; \Phi^{-1}\!\Bigl(\frac{1+\beta}{2}\Bigr),
\label{eq:CI-MIR}
\end{equation}
where $\Phi^{-1}$ is the standard normal quantile function. In the same case, an asymptotic $100\beta\%$ confidence interval for $\gamma_n$ is
\[
\widehat\gamma_n
\;\pm\;
c_\beta\,
\frac{\sqrt{2}}{n^{(1-\widehat\gamma_n)/2}\log n}.
\]
If the confidence interval for $\delta_n$ lies strictly below unity, then the asset price process is inconsistent with explosive behavior. More generally, a confidence interval for $\gamma_n$ lying well below unity indicates that the autoregressive root remains close to the non-explosive region, whereas a wide interval signals substantial uncertainty about the local persistence of the process.

\noindent When $\widehat\delta_n>1$, an asymptotic $100\beta\%$ confidence interval for $\delta_n$ is
\begin{equation}
\widehat\delta_n \;\pm\; c^{\mathcal C}_{\beta}\,\frac{2}{n^{\widehat\gamma_n}\widehat\delta_n^{\,n}},
\qquad
c^{\mathcal C}_{\beta} \;=\; \Phi_{\mathcal C}^{-1}\!\Bigl(\frac{1+\beta}{2}\Bigr),
\label{eq:CI-MER}
\end{equation}
where $\Phi_{\mathcal C}^{-1}$ denotes the two-sided quantile function of the standard Cauchy distribution. An asymptotic $100\beta\%$ confidence interval for $\gamma_n$ is then
\[
\widehat\gamma_n
\;\pm\;
c^{\mathcal C}_{\beta}\,
\frac{2}{\bigl(1+n^{-\widehat\gamma_n}\bigr)^n\log n}.
\]
If the confidence interval for $\delta_n$ lies strictly above unity, the data provide asymptotic evidence in favor of an explosive autoregressive root. Similarly in this case, the associated confidence interval for $\gamma_n$ lying well below unity suggests significantly speculative bubble in asset price. In conclusion, $\delta_n$ and $\gamma_n$  provide a natural inferential basis for distinguishing non-explosive, weakly explosive, and strongly explosive regimes.

\subsection{Date-Stamping Exuberance in Asset Price}

We follow the recursive dating methodology of \cite{phillips2011explosive}, which we view as the leading framework currently available for the econometric detection of price exuberance. Our objective is to retain the PWY dating logic while extending its calibration to environments with  time-varying volatility dynamics. To model a bubble episode explicitly, we allow the autoregressive root to be mildly explosive over a subinterval of the sample and unit-root elsewhere. Mathematically,
\begin{equation}\label{eq:mainmodel}
X_t
=
X_{t-1}\mathbf{1}\{t<\tau_e \ \text{or}\ t>\tau_f\}
\;+\;
\delta_n X_{t-1}\mathbf{1}\{\tau_e \le t \le \tau_f\}
\;+\;
\sigma_t \,\varepsilon_t,
\qquad t=1,\dots,n,
\end{equation}
where,
\begin{equation*}
\delta_n
=
1+\frac{c}{n^\alpha},
\qquad c>0,\quad \alpha\in(0,1),
\label{eq:delta-me}
\end{equation*}
and\ $\varepsilon_t \stackrel{\text{i.i.d.}}{\sim}\mathcal{N}(0,1),$ and $\sigma_t$ follows the Assumptions~\ref{ass:LLNscale}--\ref{ass:sigmahat} as explained in Section~\ref{sec:section3}.  Thus, on the interval $[\tau_e,\tau_f]$, the root satisfies $\delta_n>1$ and induces mild explosiveness, whereas outside this interval the process evolves as a unit root. We assume a standard initialization at $t=0$, for example $X_0=o_p(\sqrt{n m_{n,n}})$. Extensions with additional regressors or augmented dynamics may be handled similarly under regularity conditions analogous to those in \citet{phillips1987towards}.

\medskip
\noindent Let $\tau=[nr]$ and $\tau_0=[nr_0]$, with $0<r_0<r\le 1$. For each recursive subsample $\{X_t:t=1,\dots,\tau\}$, we estimate the corresponding autoregression and compute right-tailed SV-ADF statistics, either coefficient-based or $t$-type, along the recursion. The estimated origination date $\hat\tau_e=[n\hat r_e]$ is defined as the first upward crossing of the relevant critical boundary:
\begin{equation}\label{eq:origination}
\hat r_e
=
\inf_{\,s\in[r_0,1]}
\bigl\{\,s: \text{SV-ADF}_s^{\delta}>cv_{\beta_n,r}^{\delta}(s)\,\bigr\},
\,\,\,\,\text{or}\,\,\,\,
\hat r_e
=
\inf_{\,s\in[r_0,1]}
\bigl\{\,s: \text{SV-ADF}_s^{\,t}>cv_{\beta_n,r}^{\,t}(s)\,\bigr\},
\end{equation}
where $cv_{\beta_n,r}^{\delta}(s)$ and $cv_{\beta_n,r}^{\,t}(s)$ denote the right-tail $100(1-\beta_n)\%$ critical values for the corresponding recursive statistics under $H_0$ (unit root process) 
based on $\tau=[ns]$ observations. Since we are interested in estimating $r_e$ consistently, We allow $\beta_n\to 0$ as $n\to\infty$, so that the critical values diverge and false detections under the null become asymptotically negligible.

Conditional on an estimated origination date $\hat\tau_e$, the collapse date $\hat\tau_f=[n\hat r_f]$ is defined as the first downward crossing below the critical boundary after a short exclusion band:
\begin{equation}
\hat r_f
=
\inf_{\,s\in[\hat r_e+\log(n)/n,\,1]}
\bigl\{\,s: \text{SV-ADF}_s^{\delta}<cv_{\beta_n,l}^{\delta}(s)\,\bigr\}
\,\,\text{or}\,\,
\hat r_f
=
\inf_{\,s\in[\hat r_e+\log(n)/n,\,1]}
\bigl\{\,s: \text{SV-ADF}_s^{\,t}<cv_{\beta_n,l}^{\,t}(s)\,\bigr\}.
\label{eq:collapse}
\end{equation}
Here,  $cv_{\beta_n,l}^{\delta}(s)$ and $cv_{\beta_n,l}^{\,t}(s)$ denote the left-tail $100 \beta_n\%$ critical values for the corresponding recursive statistics under model~\eqref{eq:mainmodel} based on $\tau=[ns]$ observations. The buffer term $\log(n)/n$ rules out collapse detections occurring only $o(\log n)$ periods after the estimated origination date, thereby enforcing a minimal separation that improves numerical stability. We assume that the true dates satisfy $\tau_e=[nr_e]$ and $\tau_f=[nr_f]$ with $r_e<r_f$, so that origination and collapse are separated by an $O(n)$ interval asymptotically. The recursive rule in \eqref{eq:collapse} then compares $\text{SV-ADF}_s^{\delta}$ and $\text{SV-ADF}_s^{\,t}$ with their corresponding critical values, yielding consistent date-stamps under mildly explosive alternatives while controlling spurious detections under the null.

\medskip
\noindent

\noindent\textbf{Remark 1.} A central distinction from the conventional PWY framework lies in the calibration of the recursive critical values. In environments with time-varying volatility, thresholds derived under homoskedasticity need not be correctly aligned with the finite-sample behavior of the recursive statistics; see \cite{phillips2011dating}, and may therefore generate spurious origination signals in high-volatility periods.

\medskip
\noindent\textbf{Remark 2.} A second important distinction is that bubble origination and bubble collapse should not be governed by the same decision threshold. For origination, calibration under the unit-root null is appropriate because the pre-bubble regime is non-explosive. For collapse, however, the process has already passed through an explosive phase, so reusing the same null-based cutoff is no longer theoretically well founded. Our SV-ADF procedure therefore calibrates the two margins $cv_{\beta_n,l} (.)$ and $ cv_{\beta_n,r}(.)$ separately, allowing the origination rule and the collapse rule to reflect their different underlying distributions. This asymmetry is central for accurate date-stamping and represents an important improvement over the conventional PWY practice.

\noindent Another major drawback of PWY framework is that it considers the following specification with an explicit collapse and re-initialization mechanism. For $t=1,\dots,n,$
\begin{equation}\label{eq:reinit}
X_t
=
X_{t-1}\mathbf{1}\{t<\tau_e\}
\;+\;
\delta_n X_{t-1}\mathbf{1}\{\tau_e \le t \le \tau_f\}
\;+\;
\left(
\sum_{k=\tau_f+1}^{t}\varepsilon_k + X_{\tau_f}^{c}
\right)\mathbf{1}\{t>\tau_f\} \,+ \,\,\varepsilon_t \mathbf{1}\{t\leq\tau_f\}.
\end{equation}
Model~\eqref{eq:reinit} begins as a unit root process, becomes mildly explosive at \(\tau_e\), and returns to unit root behavior after \(\tau_f\). At collapse, the process is re-initialized at a new level \(X_{\tau_f}^{c}\), often taken to satisfy
\[
X_{\tau_f}^{c}=X_{\tau_e}+X^{c},
\qquad X^{c}=O_p(1).
\]
This formulation is theoretically convenient, since it delivers a clean post-collapse regime and simplifies asymptotic analysis. In practical terms, however, it is difficult to justify. Under mild explosiveness on \([\tau_e,\tau_f]\), the terminal value typically grows at the rate
\[
X_{\tau_f}\asymp \delta_n^{\,\tau_f-\tau_e}X_{\tau_e},
\qquad \tau_f-\tau_e=n(r_f-r_e),
\]
up to stochastic scaling, the factor \(\delta_n^{\,\tau_f-\tau_e}\) is generically exponential in the bubble span. The restriction \(X_{\tau_f}^{c}=X_{\tau_e}+O_p(1)\) therefore imposes an unrealistically sharp collapse, forcing the post-bubble level back to an \(O_p(1)\) neighborhood of the pre-bubble level despite the potentially very large magnitudes reached during the explosive phase.

For this reason, our asymptotic framework does not rely on this re-initialization assumption. Rather than viewing it as a universal feature of bubble collapse, we regard it as one particular modeling device that may be useful in some applications but is not generally implied by mildly explosive dynamics. In the \hyperref[sec:pwy_contra]{Appendix A.6}, we complement this theoretical point with simulation evidence showing that such an \(O_p(1)\) reset is implausible to reconcile with the magnitudes generated in finite samples under a mildly explosive episode.
The same recursive framework can also be applied to other unit root tests, such as Phillips--Perron tests \citep{phillips1987towards, phillips1988testing}, with SV-consistent calibration implemented in an analogous way.


\begin{theorem}
\label{thm:no-spurious}
Suppose Model~\eqref{eq:mainmodel} holds with $c=0$, i.e. no explosive episode, and let the test size satisfy $\beta_n\to 0$ as $n\to\infty$, so that the right‑tail critical values $cv_{\beta_n,r}^{\delta}$ and $cv_{\beta_n,r}^{\mathrm{t}}$ diverge. Then
\[
\Pr\!\big(\,\widehat r_e \in [r_0,1]\,\big)\ \longrightarrow\ 0.
\]
\end{theorem}

\noindent An immediate consequence of Theorem~\ref{thm:no-spurious} is that as the sample grows, the forward recursion does not falsely declare an origination date under $H_0$.  The probability of a spurious bubble origination converges to 0. 

\newcommand{\E}{\ensuremath{\mathbb{E}}}

\begin{lemma}\label{lem:main}
Let Model~\eqref{eq:mainmodel} hold with a bubble window $[\tau_e,\tau_f]$ and suppose $r\in(0,1)$ satisfies $r<r_e$, so that $t\le \lfloor nr\rfloor$ lies strictly before the origination date. Assume Assumptions~\ref{ass:LLNscale}–\ref{ass:smooth} hold. Then,
\[
\frac{X_{\lfloor nr\rfloor}}{\sqrt{n\,m_{n,n}}}\ \Rightarrow\ B(r),
\]
where $B(\cdot)$ is standard Brownian motion on $[0,1]$.
\end{lemma}
We now characterize the behavior of $\widehat r_e$ under the alternative in which a mildly explosive episode begins at $\tau_e=[nr_e]$ with $r_e>r_0$. In this case, the forward SV-ADF paths of the generating mechanism described in model \eqref{eq:mainmodel} will cross the  thresholds with probability tending to one after $r_e$, enabling consistent dating of origination and, subsequently, the collapse. 
\begin{theorem}\label{thm:orig-consistency}
Let Model~\eqref{eq:mainmodel} holds and  $\widehat r_e$ be the forward‑recursive date based on either the coefficient or the $t$‑type \text{SV-ADF} statistic, with right‑tail critical values $cv_{\beta_n,r}^{\delta}(\cdot)$ and $cv_{\beta_n,r}^{\mathrm{t}}(\cdot)$ and satisfy $\beta_n\to 0$ as $n\to\infty$. Then:
\begin{enumerate} 
\item If
\begin{equation}
\frac{1}{cv_{\beta_n,r}^{\delta}} \;+\; \frac{cv_{\beta_n,r}^{\delta}}{n^{\,1-\alpha}} \ \longrightarrow\ 0,
\label{eq:cv_cond_delta}
\end{equation}
then
\[
\widehat r_e \ \xrightarrow{p}\ r_e
\quad \text{as } n\to\infty,
\]
when $\widehat r_e$ is obtained from the coefficient based statistic $\text{SV-ADF}_r^{\delta}$.

\item If
\begin{equation}
\frac{1}{cv_{\beta_n,r}^{\mathrm{t}}} \;+\; \frac{cv_{\beta_n,r}^{\mathrm{t}}}{n^{\,1-\alpha/2}} \ \longrightarrow\ 0,
\label{eq:cv_cond_df}
\end{equation}
then
\[
\widehat r_e \ \xrightarrow{p}\ r_e
\quad \text{as } n\to\infty,
\]
when $\widehat r_e$ is obtained from the $t$‑type statistic $\text{SV-ADF}_r^{\,t}$.
\end{enumerate}
\end{theorem}

\noindent\textbf{Remark 3.} The coefficient-based SV-ADF statistic diverges at rate $n^{1-\alpha}$, while the $t$-type statistic diverges at rate $n^{1-\alpha/2}$. The associated recursive critical values remain valid under time-varying volatility and are nuisance-parameter free. As a result, the procedure remains directly implementable in practice, for example with critical values that diverge slower than polynomial rates, without requiring knowledge of $\alpha$ or separate estimation of volatility parameters.


\begin{theorem}\label{thm:collapse-consistency}
Let Model~\eqref{eq:mainmodel} holds and  $\widehat r_f$ be the forward‑recursive date based on either the coefficient or the $t$‑type SV-ADF statistic, with left‑tail critical values $cv_{\beta_n,l}^{\delta}(\cdot)$ and $cv_{\beta_n,l}^{\mathrm{t}}(\cdot)$ and satisfy $\beta_n\to 0$ as $n\to\infty$. Then conditional of $\widehat r_e >r_0$,
\begin{enumerate} 
\item If
\begin{equation}
\frac{1}{cv_{\beta_n,l}^{\delta}} \;+\; \frac{cv_{\beta_n,l}^{\delta}}{n^{\,1-\alpha}} \ \longrightarrow\ 0,
\label{eq:cv_cond_delta}
\end{equation}
then
\[
\widehat r_f \ \xrightarrow{p}\ r_f
\quad \text{as } n\to\infty,
\]
when $\widehat r_f$ is obtained from the coefficient based statistic $\text{SV-ADF}_r^{\delta}$.

\item If
\begin{equation}
\frac{n^{\alpha -1/2}}{cv_{\beta_n,l}^{\mathrm{t}}} \;+\; \frac{cv_{\beta_n,l}^{\mathrm{t}}}{n^{\,1-\alpha/2}} \ \longrightarrow\ 0,
\label{eq:cv_cond_df}
\end{equation}
then
\[
\widehat r_f \ \xrightarrow{p}\ r_f
\quad \text{as } n\to\infty,
\]
when $\widehat r_f$ is obtained from the $t$‑type statistic $\text{SV-ADF}_r^{\,t}$.
\end{enumerate}
\end{theorem}

\noindent\textbf{Remark 4.} The key departure of our asymptotic results from PWY is because we do not impose a restrictive collapse-and-reset specification after the explosive phase, and allow more general post-bubble dynamics. The coefficient-based recursive statistic is the more useful in practical empirical procedures. Its critical sequence \(cv_{\beta_n,l}^{\delta}(\cdot)\) remains nuisance-parameter free, so implementation may proceed with slowly diverging thresholds, such as \(\log n\), without requiring knowledge of \(\alpha\). By contrast, calibration of the \(t\)-type critical values \(cv_{\beta_n,l}^{\mathrm{t}}(\cdot)\) depends explicitly on \(\alpha\), which is unknown in practice. Although one could estimate \(\alpha\) and then tune the critical values accordingly, this approach is less reliable, since the resulting estimator may fall outside the relevant interval \((0,1)\) \citep{phillips2023estimation}. For this reason, we regard the coefficient-based recursive statistic as the more robust and practically implementable procedure.




\section{Real-data Experiments}\label{sec:simulation}

\newcommand{\bubble}{\textcolor{green!60!black}{\ding{51}}}
\newcommand{\nobubble}{\textcolor{red!75!black}{\ding{55}}}
\newcommand{\na}{--}

We apply our SV-ADF procedure to date-stamp bubble origination and collapse for the Magnificent Seven and a set of major AI-semiconductor firms over 2018--2026. Throughout, we compare the resulting date-stamps with those obtained from the standard PWY approach in order to assess which method provides a more credible and empirically informative detection of bubble episodes. We do not benchmark with the PSY procedure, since PSY is designed for multiple-bubble environments, whereas our framework, by construction, supports a single bubble episode for each stock.


Because the analysis is conducted at the daily frequency, transient threshold crossings need not reflect economically meaningful exuberance. To filter out short-lived fluctuations, we therefore define bubble origination as the first date at which the recursive SV-ADF statistic exceeds the origination threshold for at least two consecutive calendar months, allowing for a brief consolidation period, and bubble collapse as the first date at which the statistic falls below the collapse threshold and remains below it for one month. The rigorous construction of the thresholds is presented in Section~\ref{sec:thresholds}. Although the analysis covers 12 firms over 2018--2026, the main concentration of detected exuberance occurs in the post-2022 AI expansion.

\begin{figure}[p]
\centering
\includegraphics[
  width=\linewidth,
  height=\textheight,
  keepaspectratio,
  trim={0 0 0 0},
  clip
]{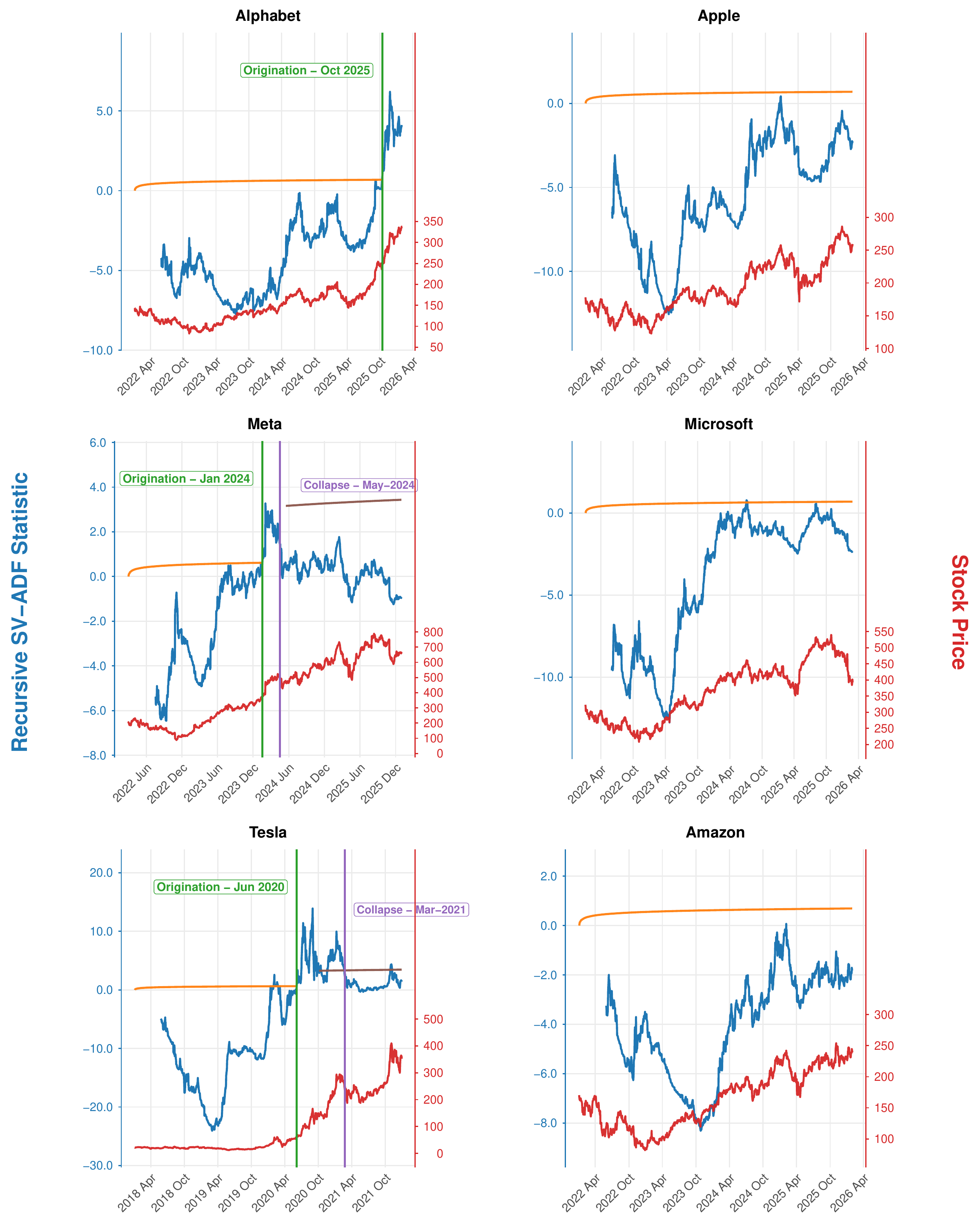}
\captionsetup{skip=0pt}
\caption{\footnotesize Recursive SV-ADF statistic (blue) and daily stock price (red) for the six largest technology firms by market capitalization. Plotting both series provides visual inspection on whether threshold crossings in the recursive statistic coincide with meaningful price movements. The orange line denotes the 90\% upper critical value under $H_0$ used for bubble origination, while the brown line denotes the 10\% lower critical value under $H_1$ used for bubble collapse. Green and violet vertical lines respectively mark the estimated origination and collapse dates from the SV-ADF procedure. The sample period is 2022--2026, except for Tesla, 2018--2022.}
\label{fig:sixtechstocks}
\end{figure}

Figure~\ref{fig:sixtechstocks} date-stamps the bubble episodes for 6 major Tech stocks. The most prominent finding is an ongoing bubble in Alphabet, which SV-ADF dates from October 2025 and which remains unresolved at the end of the sample. This episode is consistent with a sustained upward revision in investor expectations as Alphabet became increasingly viewed as a major AI beneficiary, supported in part by stronger Google Cloud performance. Tesla shows a milder but still significant episode, with bubble dynamics dated from June 2020 to March 2021, broadly consistent with a sequence of reinforcing catalysts including profitability confirmation, the stock split, and S\&P 500 inclusion. Meta also exhibits a short-lived episode, but its limited duration makes it less central to the broader pattern. For Microsoft, Apple, and Amazon, the evidence does not indicate a sustained exuberant regime.

\begin{figure}[p]
\centering
\includegraphics[
  width=\linewidth,
  height=\textheight,
  keepaspectratio,
  trim={0 0 0 0},
  clip
]{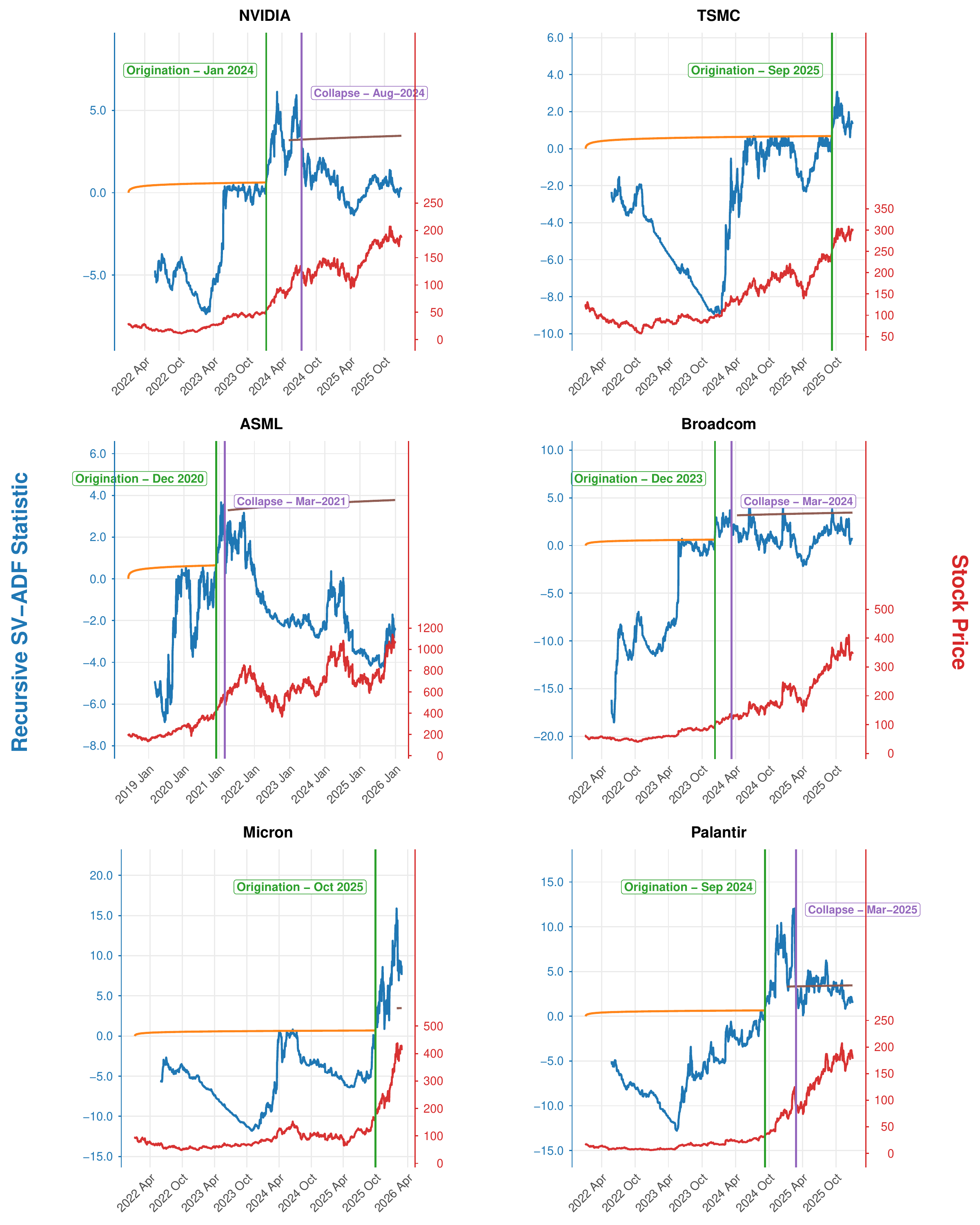}
\captionsetup{skip=0pt}
\caption{\footnotesize Recursive SV-ADF Statistic (in blue) and daily stock price (in red) for 6 largest semiconductor/AI-infrastructure firms by market capitalization. The sample period is 2022--2026, except ASML July 2018--2026. The color scheme is the same as Figure~\ref{fig:sixtechstocks}.}
\label{fig:chipai6}
\end{figure}

Figure~\ref{fig:chipai6} date-stamps bubble episodes for six major semiconductor and AI-infrastructure firms. Nvidia provides the strongest evidence: the bubble dated from November 2023 to September 2024 aligns closely with the news chronology, beginning after the November 2023 earnings surprise and intensifying following the February 2024 release. Its collapse in September 2024 is likewise consistent with a moderation in upward revisions after the August 2024 earnings announcement, despite still-strong operating performance. TSMC’s bubble begins in September 2025 and remains ongoing, reflecting its bottleneck role in advanced AI chips, while Micron’s bubble begins in October 2025 and remains ongoing with even greater intensity as high-bandwidth memory is re-rated from a cyclical product to a structurally scarce AI input. Palantir, Broadcom, and ASML also display shorter-lived and less persistent bubble-like episodes, suggesting more temporary repricing during periods of heightened AI and semiconductor enthusiasm.

\begin{figure}[!htpb]
\centering
\includegraphics[width=\linewidth]{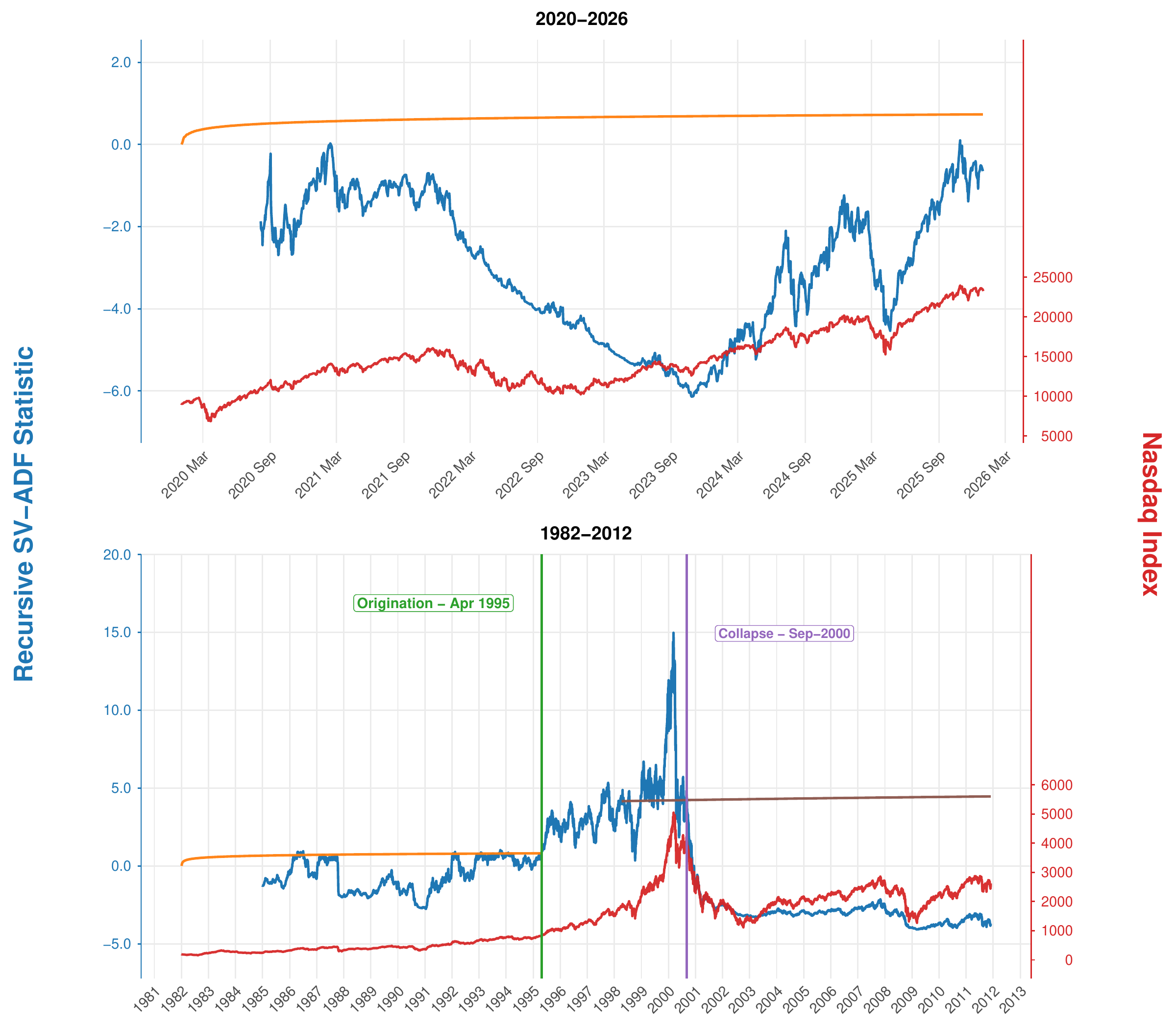}  
\captionsetup{skip=0pt}
\caption{\footnotesize: Date-stamping bubble behaviour in Nasdaq index. Time period for upper panel: 2020--2026, no exuberance detected. Time period for lower panel: 1982--2012, significant exuberance detected in 1995-2000. Date-stamps are accurate and provide early signals as compared to PWY. The color scheme is the same same as Figure~\ref{fig:sixtechstocks}.} 
\label{fig:nasdaq2} 
\end{figure}

\newpage
We replicate the bubble identification exercise in 1990s Nasdaq using a sample window that mirrors  \cite{phillips2011explosive}. Despite more robust thresholds, our SV‑ADF dates origination to early April 1995 and collapse to September 2000 (see lower panel of Figure~\ref{fig:nasdaq2}), versus June 1995 to September 2000 in PWY approach. The earlier signal with stricter control underscores the gain from volatility‑robust inference for real‑time monitoring. The upper panel of Figure \ref{fig:nasdaq2} confirms no statistically significant exuberance in the Nasdaq Index over 2020–2026, despite the AI‑driven period.

\begin{figure}[!htb]
\centering
\includegraphics[width=\linewidth]{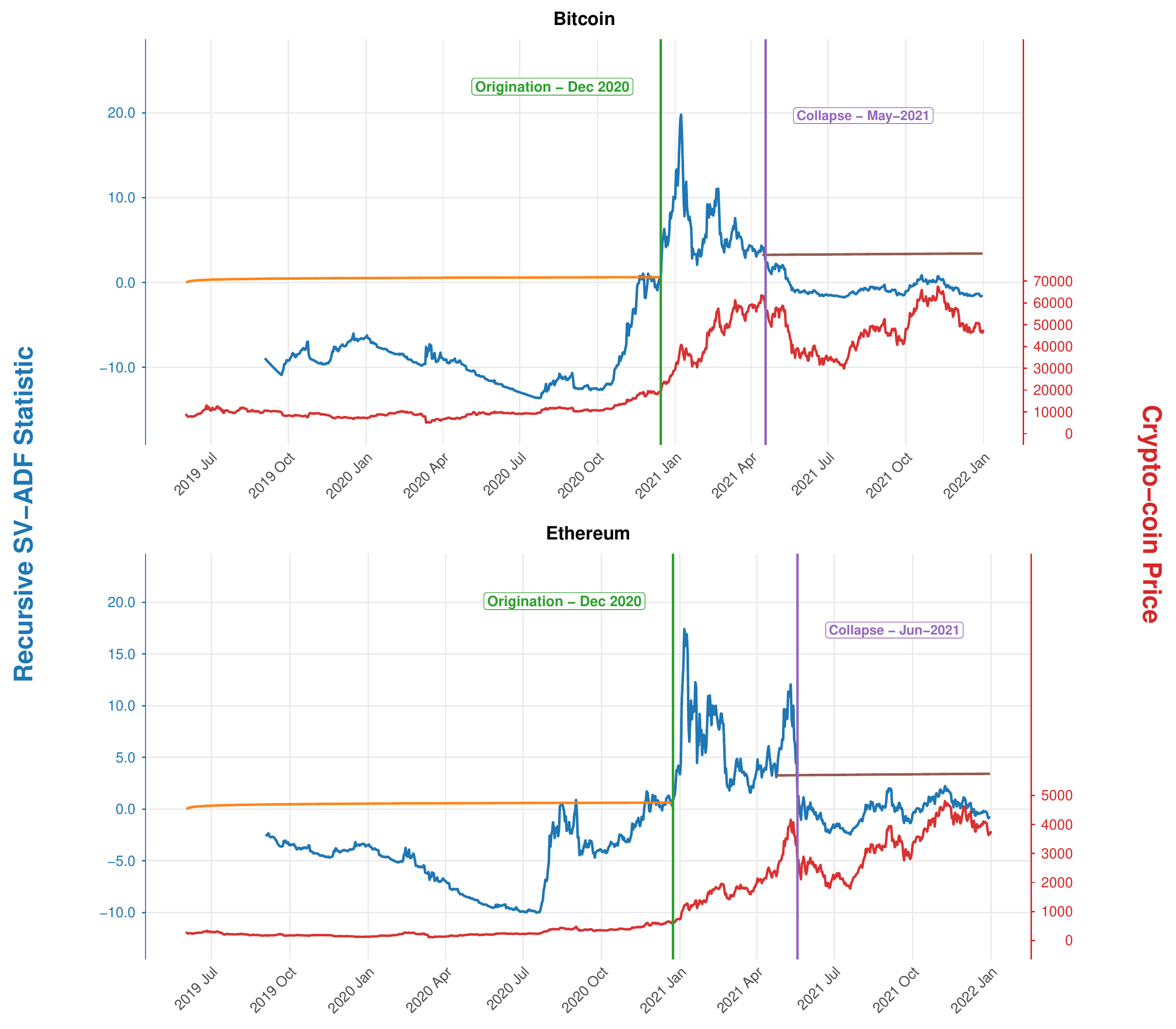}   
\captionsetup{skip=0pt}
\caption{\footnotesize  Recursive SV-ADF test suggests significant bubble in Bitcoin and Ethereum during January 2021 -- June 2021. Similar cryptocurrencies during this era exhibited pronounced exuberance (see the \hyperref[sec:appendix]{Appendix}). The color scheme is the same as Figure~\ref{fig:sixtechstocks}.}
\label{fig:crypto}
\end{figure}

Next, we apply our date-stamping procedure to Bitcoin and Ethereum. The resulting episodes in Figure~\ref{fig:crypto} are broadly consistent with the bubble patterns identified by devolatilized recursive tests by \cite{boswijk2024testing}. However, the standard PWY procedure under homoskedasticity fails to detect accurate date-stamps in cryptocurrencies. This difference is striking because properly accounting for volatility makes statistically significant bubble behavior detectable in major cryptoassets (see \hyperref[sec:more_crypto]{Appendix A.7}).

\begin{table}[!htpb]
\centering
\scriptsize
\setlength{\tabcolsep}{4pt}
\begin{tabular}{l cc @{\hspace{1.5em}} cc}
\toprule
& \multicolumn{2}{c}{2018--2020} & \multicolumn{2}{c}{2022--2026} \\
\cmidrule(r){2-3}\cmidrule(l){4-5}
Company & SV-ADF & PWY & SV-ADF & PWY \\
\midrule
Apple      & \nobubble & \bubble   & \nobubble & \bubble   \\
Microsoft  & \nobubble & \bubble   & \nobubble & \bubble   \\
Nvidia     & \nobubble & \bubble   & \bubble   & \bubble   \\
Alphabet   & \nobubble & \bubble   & \bubble   & \bubble   \\
Amazon     & \nobubble & \bubble   & \nobubble & \bubble   \\
Meta       & \nobubble & \bubble   & \nobubble & \bubble   \\
Tesla      & \bubble   & \bubble   & \nobubble & \nobubble \\
TSMC       & \nobubble & \bubble   & \bubble   & \bubble   \\
Broadcom   & \nobubble & \bubble   & \bubble   & \bubble   \\
Palantir   & \na       & \na       & \bubble   & \bubble   \\
ASML       & \nobubble & \bubble   & \nobubble & \bubble   \\
Micron     & \nobubble & \nobubble & \bubble   & \bubble   \\
\bottomrule
\end{tabular}
\captionsetup{skip=0pt}
\caption{\footnotesize In post-2022 AI-driven era, SV-ADF procedure identifies bubble in Alphabet and almost every semiconductors, while the identified bubble in 2018--2020 is only in Tesla. However, PWY incorrectly classifies nearly the entire sample as bubbly in most periods. Palantir went public in September 2020 and is not meaningfully testable within the 2018--2020 subsample.}
\label{tab:bubble_decisions_two_periods}
\end{table}

Beyond date-stamping bubble episodes, we directly test for the presence of exuberance in the post-2022 AI period at the level of individual stocks. We estimate the autoregressive coefficient $\widehat \delta_n$ from \eqref{eq:main_model_used} and classify a stock as exuberant whenever the associated SV-ADF statistic exceeds its theoretical right-tail critical value, which is close to the 90th-percentile upper quantile under the null. Table~\ref{tab:bubble_decisions_two_periods} compares these classifications with those implied by the standard PWY procedure. The contrast is sharp, PWY rejects the null for nearly all stocks, for any time period considered, effectively labeling almost the entire set as bubbly over the sample, that is difficult to regard as economically plausible. By contrast, our SV-ADF approach is far more selective and confirms pronounced bubble behavior only for Tesla in 2018–2020 and, in the post-2022 AI period, only for Alphabet among the large technology firms, while indicating notable speculative behavior across much of the semiconductor and AI-infrastructure segment.

\begin{figure}[!htpb]
\centering
\includegraphics[width=0.8\linewidth]{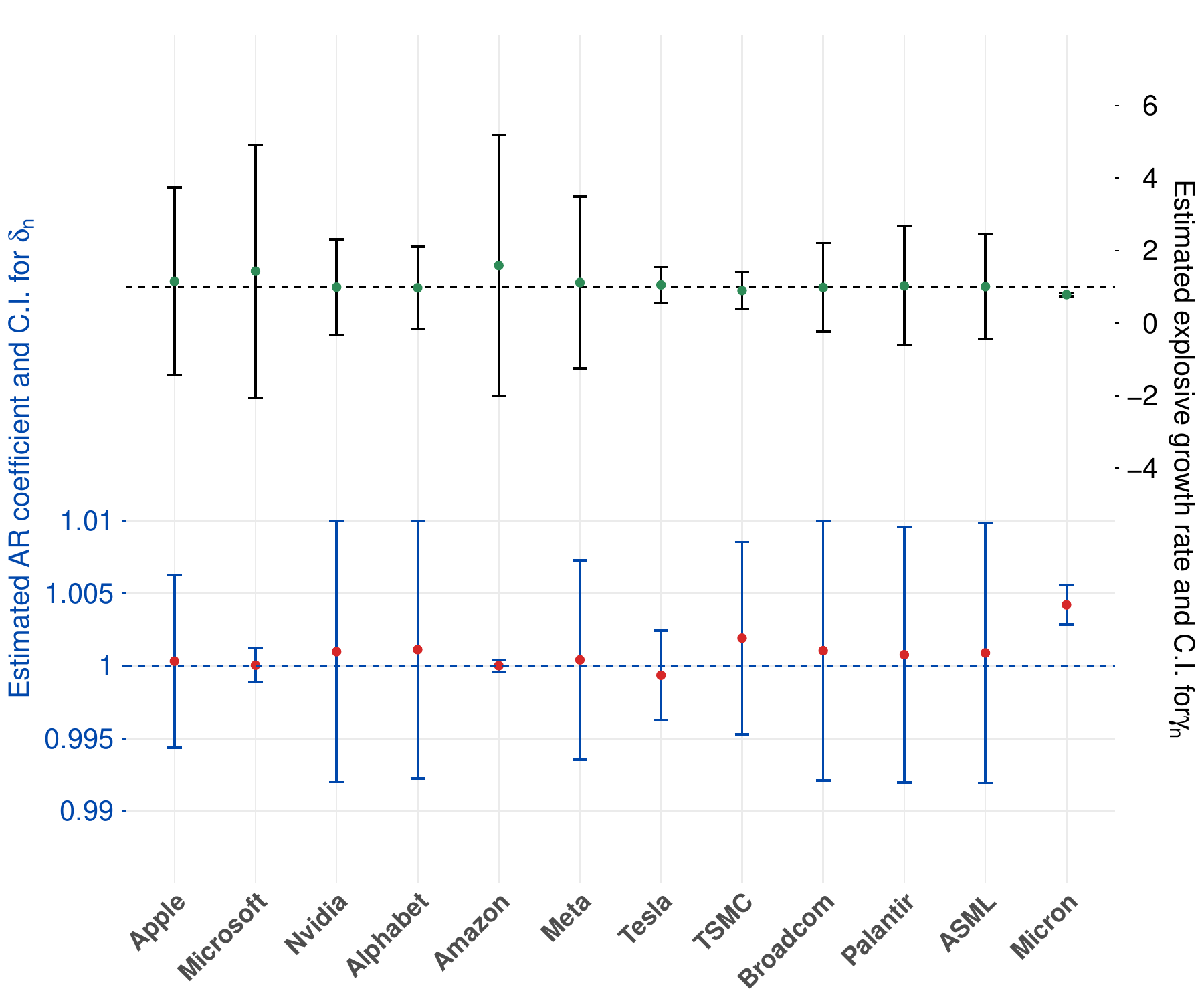} 
\captionsetup{skip=0pt}
\caption{\footnotesize  Estimated autoregressive coefficients, growth rates, and their 95\% C.I over 2022--2026. For most firms, unity lies within the interval for the AR coefficient, consistent with near-unit-root behavior.}
\label{fig:12stocks_ar} 
\end{figure}

Over the 2022–2026 post-AI sample, we also report 95\% confidence intervals for the autoregressive coefficient and the explosive growth rate as suggested in \citep{sarkar2025double, phillips2023estimation}. A narrow interval that excludes unity indicates a process that is materially away from a unit root. In Figure~\ref{fig:12stocks_ar}, all firms except Tesla has estimated coefficient more than 1, suggesting explosive behavior.  Tesla has $\hat \delta < 1$, suggesting the weakest evidence of exuberance in this recent AI boom.  Micron is the notable exception since its interval lies entirely above unity, placing it clearly apart from the rest of the sample. Its estimated growth rate C.I. is entirely below 1 too, providing especially strong evidence of bubble behavior, consistent with the SV-ADF date-stamping results.

\subsection{Threshold Selection Insights}\label{sec:thresholds}
In this section, we discuss the calibration of the cross-validated thresholds used in the SV-ADF tests and, in turn, the decision rules for accepting or rejecting the hypothesis involving bubble detection. We simulate data from model~\eqref{eq:main_model_used} under \(H_0\) ($\delta=1$), using the volatility specifications in \cite{sarkar2025double}. For each sample size \(n \in \{100,200,\dots,1000\}\), we generate 1000 independent replications and compute the 90th percentile of the resulting SV-ADF statistic. Guided by both the calibrated values and the asymptotic implications of Theorems~\ref{thm:orig-consistency} and~\ref{thm:collapse-consistency}, we find that these upper 90 percent critical values are well approximated by \(\log(n)/10\), which is adopted as the origination threshold for speculative bubbles.

\begin{figure}[!htpb]
\centering
\includegraphics[width=\linewidth]{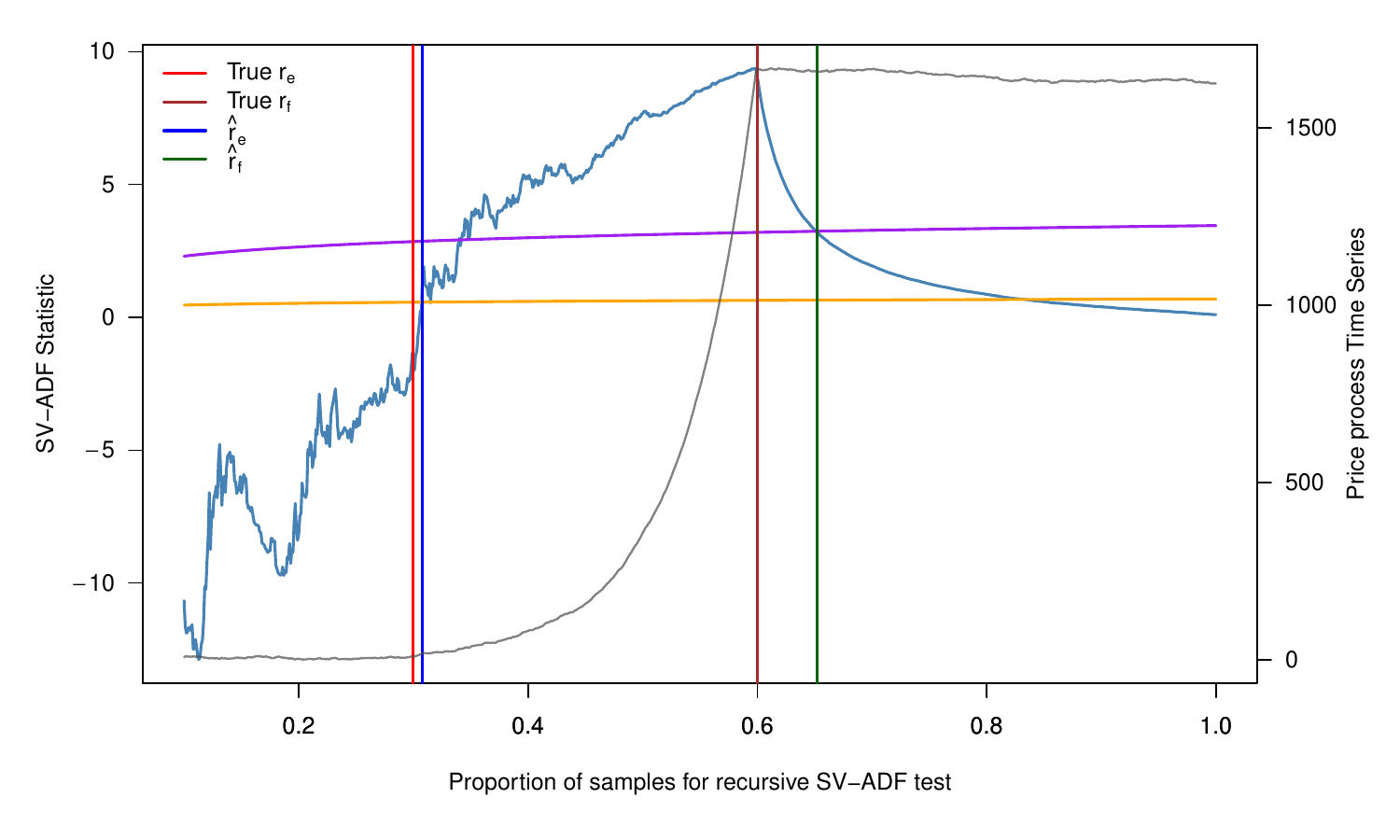}  
\captionsetup{skip=0pt}
\caption{\footnotesize  One simulated time series path (in black) and recursive SV-ADF statistic (in blue). The red and brown vertical lines denote the true $r_e =0.3, r_f=0.6$ respectively, while the blue and green mark the estimated $\hat r_e =0.308, \hat r_f=0.652$. The orange horizontal line is the origination threshold ($\log(ns)/10$), purple line is collapse threshold ($\log(ns)/2$). \cite{phillips2011explosive} considered a uniform threshold of $\log \log (ns)/100$ for both origination and collapse. }
\label{fig:one_simulation}
\end{figure}

For bubble collapse, the relevant threshold is based on the lower 10th percentile of the ADF statistic under \(H_1\). Because this distribution  under \(H_1\) depends on nuisance parameters in the model, and hence the threshold may vary across specifications, we proceed differently: for each \(n\), we randomly draw values of \(r_e, r_f, c, d, \alpha\), and $\eta,$ compute the corresponding 10th percentile threshold of the SV-ADF statistic, and then average across draws. The resulting sequence is most closely approximated by \(\log(n)/2\), which we therefore use as the collapse threshold.  The code for the calibration exercise, along with more detailed discussion on thresholds choices and robustness checks are provided in \hyperref[sec:threshs]{Appendix A.8} and
\href{https://github.com/AbirS2026/SVADF-Bubble}{GitHub}. Figure~\ref{fig:one_simulation} below provides a visual representation of the estimated start and end points of a bubble for one simulated time series path.

As explained in Section~\ref{sec:section4}, the conventional PWY procedure applies the same threshold; see Table 1 in \cite{phillips2011explosive}, to both bubble origination and collapse. Our procedure instead calibrates the origination and collapse thresholds separately, allowing the two margins of the episode to reflect different underlying distributions. Hence, it is expected to deliver more accurate date-stamps for bubble origination and collapse, together with improved power in the corresponding statistical procedures. Table~\ref{tab:id_rates} reports bubble origination and collapse identification rates across alternative simulated configurations of \((r_e,r_f)\), comparing the proposed SV-ADF procedure with the PWY approach. These results show that the SV-ADF procedure delivers uniformly higher power to detect both bubble origination and collapse across the regimes considered. 

\begin{table}[!htpb]
\centering
\scriptsize
\setlength{\tabcolsep}{4pt}
\begin{tabular}{c c @{\hspace{1.2em}} r r @{\hspace{0.8em}} r r}
\toprule
& & \multicolumn{2}{c}{SV-ADF} & \multicolumn{2}{c}{PWY} \\
\cmidrule(r){3-4}\cmidrule(l){5-6}
$r_e$ & $r_f$ & Origination & Collapse & Origination & Collapse \\
\midrule
0.2 & 0.50 & 0.976 & 0.953 & 0.949 & 0.919 \\
0.2 & 0.65 & 0.971 & 0.950 & 0.954 & 0.930 \\
0.2 & 0.75 & 0.976 & 0.962 & 0.963 & 0.944 \\
0.3 & 0.50 & 0.822 & 0.816 & 0.757 & 0.745 \\
0.3 & 0.65 & 0.782 & 0.773 & 0.714 & 0.703 \\
0.3 & 0.75 & 0.823 & 0.812 & 0.755 & 0.743 \\
0.4 & 0.65 & 0.722 & 0.716 & 0.618 & 0.609 \\
0.4 & 0.75 & 0.714 & 0.712 & 0.615 & 0.612 \\
0.5 & 0.65 & 0.636 & 0.632 & 0.532 & 0.529 \\
0.5 & 0.75 & 0.654 & 0.653 & 0.561 & 0.558 \\
\bottomrule
\end{tabular}
\captionsetup{skip=0pt}
\caption{\footnotesize Power of SV-ADF procedure is uniformly better than PWY. Our method delivers higher origination and collapse identification rates than PWY across all reported \((r_e,r_f)\) configurations. The gains in power are notable as the bubble window shifts later or becomes more compressed, since accurate timing is more demanding.}
\label{tab:id_rates}
\end{table}

We similarly compare the effective power, of the SV-ADF and PWY procedures across a range of volatility regimes. Specifically, we consider (i) the homoskedastic case, (ii) stochastic volatility under a double-local-to-unity specification, and (iii) a GARCH setting with near-nonstationary volatility \((\alpha+\beta \approx 1)\), and evaluate performance across different parameter configurations. Table~\ref{tab:id_rates_vol_9cases} reports the resulting bubble origination and collapse identification rates. Consistent with the earlier evidence, the SV-ADF procedure exhibits uniformly higher power than the PWY approach across all model designs.

\begin{table}[H]
\centering
\scriptsize
\setlength{\tabcolsep}{4pt}
\begin{tabular}{l l @{\hspace{2.0em}} r r @{\hspace{1.0em}} r r}
\toprule
& & \multicolumn{2}{c}{SV-ADF} & \multicolumn{2}{c}{PWY} \\
\cmidrule(r){3-4}\cmidrule(l){5-6}
Volatility regime & Design & Origination & Collapse & Origination & Collapse \\
\midrule
\multirow{3}{*}{\textit{Homoskedastic}}
& $c=0.3$ & 0.814 & 0.816 & 0.726 & 0.711 \\
& $c=0.5$ & 0.817 & 0.803 & 0.738 & 0.715 \\
& $c=1.0$ & 0.825 & 0.816 & 0.762 & 0.755 \\
\addlinespace
\multirow{3}{*}{\textit{AR(1) log-volatility}}
& $c=0.3,\ \eta=0.1$ & 0.795 & 0.791 & 0.738 & 0.717 \\
& $c=0.5,\ \eta=0.5$ & 0.714 & 0.693 & 0.666 & 0.647 \\
& $c=1.0,\ \eta=1.0$ & 0.659 & 0.640 & 0.624 & 0.618 \\
\addlinespace
\multirow{3}{*}{\textit{GARCH}}
& $c=0.3,\ \alpha_g=0.05,\ \beta_g=0.94$ & 0.799 & 0.781 & 0.759 & 0.732 \\
& $c=0.5,\ \alpha_g=0.10,\ \beta_g=0.89$ & 0.799 & 0.784 & 0.749 & 0.732 \\
& $c=1.0,\ \alpha_g=0.05,\ \beta_g=0.94$ & 0.825 & 0.819 & 0.758 & 0.756 \\
\bottomrule
\end{tabular}
\captionsetup{skip=0pt}
\caption{\footnotesize The SV-ADF procedure uniformly outperforms the PWY approach in both origination and collapse identification across all reported volatility regimes. For clarity in comparison, we have fixed $r_e =0.3, r_f =0.6$. For details on the nuisance parameters $c,\alpha_g,$ and $\beta_g$ in GARCH, see the \hyperref[sec:garch]{Appendix A.9}.}
\label{tab:id_rates_vol_9cases}
\end{table}

Finally, Table~\ref{tab:bias_mse_selected9} report the estimated values of \(\hat r_e\) and \(\hat r_f\), together with their mean squared errors (MSE), across alternative choices of \((r_e,r_f,\alpha)\). The results indicate that the proposed estimator exhibits uniformly low bias and MSE across the specifications considered, by showing that the gains in detection power do not come at the expense of estimation accuracy.
\begin{table}[H]
\centering
\scriptsize
\setlength{\tabcolsep}{4pt}
\begin{tabular}{c c c @{\hspace{2.0em}} c c c c @{\hspace{1.2em}} c c c c}
\toprule
& & & \multicolumn{4}{c}{SV-ADF} & \multicolumn{4}{c}{PWY} \\
\cmidrule(r){4-7}\cmidrule(l){8-11}
$r_e$ & $r_f$ & $\alpha$
& $\widehat r_e$ & $\widehat r_f$ & MSE$(\widehat r_e)$ & MSE$(\widehat r_f)$
& $\widehat r_e$ & $\widehat r_f$ & MSE$(\widehat r_e)$ & MSE$(\widehat r_f)$ \\
\midrule
0.2 & 0.50 & 0.3 & 0.2040 & 0.5100 & 0.0014 & 0.0183 & 0.1966 & 0.4290 & 0.0019 & 0.0234 \\
0.2 & 0.65 & 0.3 & 0.2012 & 0.6438 & 0.0015 & 0.0418 & 0.1943 & 0.5321 & 0.0019 & 0.0561 \\
0.3 & 0.50 & 0.3 & 0.2830 & 0.4856 & 0.0054 & 0.0223 & 0.2659 & 0.4047 & 0.0079 & 0.0286 \\
0.4 & 0.50 & 0.5 & 0.3654 & 0.4330 & 0.0230 & 0.0285 & 0.3407 & 0.4268 & 0.0308 & 0.0463 \\
0.4 & 0.65 & 0.5 & 0.3787 & 0.5447 & 0.0224 & 0.0606 & 0.3499 & 0.4587 & 0.0272 & 0.0798 \\
0.4 & 0.75 & 0.5 & 0.3914 & 0.6320 & 0.0221 & 0.0862 & 0.3596 & 0.5066 & 0.0270 & 0.1189 \\
0.5 & 0.65 & 0.7 & 0.3968 & 0.4398 & 0.0652 & 0.0952 & 0.3595 & 0.4133 & 0.0711 & 0.1096 \\
0.4 & 0.65 & 1.0 & 0.3793 & 0.4231 & 0.0590 & 0.1069 & 0.3545 & 0.3995 & 0.0602 & 0.1177 \\
\bottomrule
\end{tabular}
\captionsetup{skip=0pt}
\caption{\footnotesize Accuracy comparison of SV-ADF vs PWY. Across the reported designs, our proposed estimator delivers better $\hat r_e$ and $\hat r_f$, systematically closer to the true values and has lower MSE than the corresponding PWY estimates.}
\label{tab:bias_mse_selected9}
\end{table}

\newpage
\section{Conclusion}\label{sec:conclusion}

This paper studied bubble detection and consistent date-stamping in environments where volatility is time-varying, with particular emphasis on the recent AI-driven revaluation of large technology and semiconductor firms. The main message is that volatility dynamics are not a secondary complication in this setting, they are central to credible inference. Once volatility is allowed to evolve systematically, both the construction of recursive ADF tests and the calibration of the thresholds used for bubble origination and collapse need to be reconsidered. The SV-ADF framework developed here provides a practical way to do so while retaining tractable asymptotic theory and direct empirical implementability.

A central practical implication of the paper is that the conventional PWY benchmark appears too coarse in this high-volatility setting. In our empirical applications, PWY tends to classify an implausibly large share of assets as exuberant, particularly when volatility is elevated and persistent. The SV-ADF procedure is more selective and less prone to volatility-induced false positives, yet it continues to detect the more durable and quantitatively pronounced episodes of explosiveness. 

The empirical results point to substantial heterogeneity in speculative behavior across AI-exposed assets. Among the large technology firms, the strongest current evidence indicates that Alphabet remains in an ongoing bubble, while Tesla exhibits a clear earlier episode during 2020--2021. In the semiconductor and AI-infrastructure segment, the evidence is even stronger: Nvidia provides the sharpest validation of the procedure through a bubble window that aligns closely with major earnings-driven revisions, and both TSMC and Micron appear to be in pronounced and ongoing bubble regimes late in the sample. By contrast, several other firms display only short-lived or less persistent episodes, indicating that the recent AI cycle should not be viewed as a uniform cross-sectional bubble. At the broader market level, the Nasdaq does not exhibit statistically significant exuberance over 2020--2026. In addition, the cryptocurrency analysis identifies exuberant  episodes in Bitcoin and Ethereum during 2021, with date-stamps that align closely with recent volatility-adjusted evidence \citep{boswijk2024testing}, and further illustrate the importance of allowing for time-varying volatility in practical bubble detection.

Several directions remain open for future work. One natural extension is to allow explicitly for multiple bubble episodes within the same asset while preserving volatility-robust calibration. We intend to pursue this question in subsequent work and to compare the resulting methodology with the standard PSY benchmark \citep{phillips2015testing}, we also plan to develop an R package for bubble dating that can be leveraged for real-time monitoring. Another potential research direction is to integrate higher-frequency volatility information, options-based measures \citep{fusari2025testing}, or market microstructure adjustments \citep{zhang2005tale} into the recursive dating framework. These extensions would further strengthen the usefulness of volatility-robust bubble diagnostics in periods of rapid technological and financial change.
\newpage
\bibliography{bibliographyCiteDrive.bib} 
\appendix
\numberwithin{equation}{section} 

\makeatletter
\renewcommand\@seccntformat[1]{\csname the#1\endcsname.\quad}
\makeatother

\section{Appendix}\label{sec:appendix}
\subsection{Proof of Theorem~\ref{thm:SVADF-null}}

\noindent
We provide a proof sketch. Fix \(r\in[r_0,1]\) and set
\(\tau=\lfloor nr\rfloor\). Under \(H_0:\delta=1\),
\[
X_t=X_0+\sum_{j=1}^t u_j,
\qquad
u_j=\sigma_j\varepsilon_j,
\qquad
\varepsilon_j\stackrel{\mathrm{i.i.d.}}{\sim}N(0,1),
\]
with \(X_0=o_p(\sqrt{n\,m_{n,n}})\). Define the normalized partial-sum process
\[
W_n(s)
:=
\frac{1}{\sqrt{n\,m_{n,n}}}
\sum_{t=1}^{\lfloor ns\rfloor}u_t,
\qquad s\in[0,1].
\]
Conditional on the volatility path, \(W_n\) is a centered Gaussian process with
independent increments and conditional variance function
\[
V_n(s)
:=
\langle W_n\rangle(s)
=
\frac{1}{n\,m_{n,n}}
\sum_{t=1}^{\lfloor ns\rfloor}\sigma_t^2 .
\]
By Assumption~\ref{ass:LLNscale},
\(
\sup_{s\in[0,1]}|V_n(s)-s|\xrightarrow{p}0.
\) Therefore, conditional on the volatility path, the normalized partial-sum process
is asymptotically a Brownian motion with linear time scale. Hence
\[
W_n(\cdot)\Rightarrow W(\cdot)
\qquad\text{in }D[0,1],
\]
where \(W\) is standard Brownian motion. Now, we define the right-continuous process
\[
W_n\!\Big(\frac{t}{n}\Big)
:=
\frac{1}{\sqrt{n\,m_{n,t}}}\sum_{j=1}^{t}\sigma_j\varepsilon_j,
\qquad
W_n(u):=
W_n\!\Big(\frac{\lfloor nu\rfloor}{n}\Big),
\quad u\in[0,1].
\]
Let $
a_{n,t}:=\sqrt{\frac{m_{n,t}}{m_{n,n}}},$
then
\(
a_{n,t}W_n\!\Big(\frac{t}{n}\Big)
=
\frac{1}{\sqrt{n\,m_{n,n}}}\sum_{j=1}^{t}\sigma_j\varepsilon_j
\)
By Assumption~\ref{ass:smooth}, \(a_{n,t}=1+o(1)\) uniformly on the recursive
range, and hence \(W_n(\cdot)\Rightarrow W(\cdot)\) on \([r_0,1]\). Then,
\[
X_t
=
X_0+\sum_{j=1}^{t}\sigma_j\varepsilon_j
=
X_0+\sqrt{n\,m_{n,t}}\,W_n\!\Big(\frac{t}{n}\Big)
=
X_0+\sqrt{n\,m_{n,n}}\,
a_{n,t}W_n\!\Big(\frac{t}{n}\Big).
\]
Let
\(
\bar X_\tau
:=
\frac{1}{\tau}\sum_{t=1}^{\tau}X_{t-1}
\),
then
\[
\bar X_\tau
=
X_0+\sqrt{n\,m_{n,n}}
\left(
\frac{1}{\tau}\sum_{t=1}^{\tau}
a_{n,t-1}W_n\!\Big(\frac{t-1}{n}\Big)
\right).
\]
By the step-function representation,
\[
\frac{1}{n}\sum_{t=1}^{\tau}
a_{n,t-1}W_n\!\Big(\frac{t-1}{n}\Big)
=
\int_0^{\tau/n}
a_{n,\lfloor nu\rfloor}W_n(u)\,du+o_p(1).
\]
Since \(a_{n,\lfloor nu\rfloor}=1+o(1)\) uniformly on the recursive range,
and \(\tau/n\to r\), this gives
\[
\frac{1}{\tau}\sum_{t=1}^{\tau}
a_{n,t-1}W_n\!\Big(\frac{t-1}{n}\Big)
=
\frac{1}{r}\int_0^r W_n(u)\,du+o_p(1).
\]
Therefore,
\begin{equation}\label{eq:mean-correct}
\bar X_\tau
=
X_0+\sqrt{n\,m_{n,n}}
\left[
\frac{1}{r}\int_0^r W_n(u)\,du+o_p(1)
\right].
\end{equation}

\noindent
Demeaning yields,
\begin{align}
\frac{\widetilde X_{t-1}}{\sqrt{n\,m_{n,n}}}
&=
a_{n,t-1}W_n\!\Big(\frac{t-1}{n}\Big)
-
\frac{1}{r}\int_0^r W_n(u)\,du
+o_p(1) \notag\\
&=
r_{n,t-1}W_n\!\Big(\frac{t-1}{n}\Big)
+
\widetilde W_n\!\Big(\frac{t-1}{n}\Big)
+o_p(1),
\label{eq:Xtilde}
\end{align}
where
\[
r_{n,t}:=a_{n,t}-1,
\qquad
\widetilde W_n(v)
:=
W_n(v)-\frac{1}{r}\int_0^r W_n(u)\,du .
\]
Then \(\widetilde W_n(\cdot)\Rightarrow \widetilde W_r(\cdot)\) on \([0,r]\). From \eqref{eq:Xtilde},
\begin{align}\label{eq:Xtilde_square}
\frac{\widetilde X_{t-1}^2}{n\,m_{n,n}}
&=
r_{n,t-1}^2
W_n\!\left(\frac{t-1}{n}\right)^2
+
\widetilde W_n\!\left(\frac{t-1}{n}\right)^2 \notag\\
&\qquad
+
2r_{n,t-1}
W_n\!\left(\frac{t-1}{n}\right)
\widetilde W_n\!\left(\frac{t-1}{n}\right)
+o_p(1).
\end{align}
The terms involving \(r_{n,t-1}\) do not contribute in the sum, since
\(\sup_u|W_n(u)|=O_p(1)\), \(\sup_u|\widetilde W_n(u)|=O_p(1)\), and
\(r_{n,t}=o(1)\) uniformly on the recursive range. Hence,
\begin{align}\label{eq:Xtilde_square_main}
\frac{1}{n^2\,m_{n,n}}
\sum_{t=1}^{\tau}\widetilde X_{t-1}^2
&=
\frac{1}{n}\sum_{t=1}^{\tau}
\widetilde W_n\!\Big(\frac{t-1}{n}\Big)^2
+o_p(1) \notag\\
&\Rightarrow
\int_0^r \widetilde W_r(u)^2\,du .
\end{align}

\noindent
Similarly, using
\(
u_t
=
\sqrt{n\,m_{n,t}}W_n\!\Big(\frac{t}{n}\Big)
-
\sqrt{n\,m_{n,t-1}}W_n\!\Big(\frac{t-1}{n}\Big),
\)
we have
\begin{align*}
\frac{u_t}{\sqrt{n\,m_{n,n}}}
&=
a_{n,t}W_n\!\Big(\frac{t}{n}\Big)
-
a_{n,t-1}W_n\!\Big(\frac{t-1}{n}\Big).
\end{align*}
By Assumption~\ref{ass:smooth}, for \(t>n^\zeta\),
\[
a_{n,t}W_n\!\Big(\frac{t}{n}\Big)
-
a_{n,t-1}W_n\!\Big(\frac{t-1}{n}\Big)
=
a_{n,t}\,\Delta W_n\!\Big(\frac{t}{n}\Big)+o_p(n^{-1/2}),
\]
where
\[
\Delta W_n\!\Big(\frac{t}{n}\Big)
:=
W_n\!\Big(\frac{t}{n}\Big)
-
W_n\!\Big(\frac{t-1}{n}\Big).
\]
Therefore, applying similar argument as in \eqref{eq:Xtilde_square_main}
\begin{align}\label{eq:cross}
\sum_{t=1}^{\tau}
\frac{\widetilde X_{t-1}u_t}{n\,m_{n,n}}
&=
\sum_{t=1}^{\tau}
a_{n,t}\,\widetilde W_n\!\Big(\frac{t-1}{n}\Big)
\Delta W_n\!\Big(\frac{t}{n}\Big)
+o_p(1) \notag\\
&\Rightarrow
\int_0^r \widetilde W_r(u)\,dW(u).
\end{align}

\noindent
Combining \eqref{eq:Xtilde_square_main} and \eqref{eq:cross},
\begin{align}\label{eq:delta_stat}
\text{SV-ADF}_r^{\delta}
&=
\tau(\widehat\delta(\tau)-1) \notag
=
\tau\,
\frac{\sum_{t=1}^{\tau}\widetilde X_{t-1}u_t}
     {\sum_{t=1}^{\tau}\widetilde X_{t-1}^{2}}
\Rightarrow
r\,
\frac{\displaystyle \int_0^r \widetilde W_r(u)\,dW(u)}
     {\displaystyle \int_0^r \widetilde W_r(u)^2\,du}.
\end{align}

\noindent
By Assumption~\ref{ass:smooth}, \(m_{n,\tau}=m_{n,n}(1+o(1))\) uniformly in
\(r\), and by Assumption~\ref{ass:LLNscale},
\[
\frac{1}{\tau\,m_{n,\tau}}\sum_{t=1}^{\tau}u_t^2\xrightarrow{p}1 .
\]
\[
\frac{\sum_{t=1}^{\tau}\widetilde X_{t-1}u_t}
     {\big(m_{n,\tau}\sum_{t=1}^{\tau}\widetilde X_{t-1}^{2}\big)^{1/2}}
\Rightarrow
\frac{\displaystyle \int_0^r \widetilde W_r(u)\,dW(u)}
     {\displaystyle
     \left(\int_0^r \widetilde W_r(u)^2\,du\right)^{1/2}}.
\]

\noindent
Finally, by Assumption~\ref{ass:sigmahat},
\(
\frac{\widehat\sigma_\tau^2}{m_{n,\tau}}\xrightarrow{p}1
\).  Slutsky's theorem then yields
\begin{align}\label{eq:t_stat}
\text{SV-ADF}_r^{t}
&=
\frac{\sum_{t=1}^{\tau}\widetilde X_{t-1}u_t}
     {\big(\widehat\sigma_\tau^{2}
     \sum_{t=1}^{\tau}\widetilde X_{t-1}^{2}\big)^{1/2}}
\Rightarrow
\frac{\displaystyle \int_0^r \widetilde W_r(u)\,dW(u)}
     {\displaystyle
     \left(\int_0^r \widetilde W_r(u)^2\,du\right)^{1/2}}.
\end{align}

\subsection{Proof of Theorem~\ref{thm:no-spurious}}

\noindent
Under \(H_0:\delta_n=1\), Theorem~\ref{thm:SVADF-null} gives functional
convergence of the recursive statistics on \([r_0,1]\). Moreover, for each fixed
\(r\), the limiting null distributions do not depend on \(r\). Indeed, let
\(B_r(v)=r^{-1/2}W(rv)\), \(v\in[0,1]\). Then \(B_r\) is standard Brownian
motion and
\[
r\frac{\displaystyle\int_0^r \widetilde W_r(u)\,dW(u)}
        {\displaystyle\int_0^r \widetilde W_r(u)^2\,du}
\overset{d}{=}
\frac{\displaystyle\int_0^1 \widetilde B(v)\,dB(v)}
     {\displaystyle\int_0^1 \widetilde B(v)^2\,dv}, \qquad
\frac{\displaystyle\int_0^r \widetilde W_r(u)\,dW(u)}
     {\displaystyle\left(\int_0^r \widetilde W_r(u)^2\,du\right)^{1/2}}
\overset{d}{=}
\frac{\displaystyle\int_0^1 \widetilde B(v)\,dB(v)}
     {\displaystyle\left(\int_0^1 \widetilde B(v)^2\,dv\right)^{1/2}},
\]
where,
\[
\widetilde B(v):=B(v)-\int_0^1 B(q)\,dq .
\]
Thus the null recursive statistics are asymptotically pivotal and
\[
\sup_{r\in[r_0,1]}|\text{SV-ADF}_r^\delta|=O_p(1),
\qquad
\sup_{r\in[r_0,1]}|\text{SV-ADF}_r^t|=O_p(1).
\]

\noindent
Note,
$
c_{n,\delta}:=\inf_{s\in[r_0,1]}cv_{\beta_n,r}^\delta(s)\to\infty,$ and $c_{n,t}:=\inf_{s\in[r_0,1]}cv_{\beta_n,r}^t(s)\to\infty .$ Hence,
\[
\Pr(\widehat r_e^\delta\in[r_0,1])
\le
\Pr\left(
\sup_{r\in[r_0,1]}|\text{SV-ADF}_r^\delta|>c_{n,\delta}
\right)
\to0,
\]
and similarly,
\[
\Pr(\widehat r_e^t\in[r_0,1])
\le
\Pr\left(
\sup_{r\in[r_0,1]}|\text{SV-ADF}_r^t|>c_{n,t}
\right)
\to0.
\]
Hence, under \(H_0\), the probability of any spurious upward crossing vanishes.

\subsection{Proof of Lemma~\ref{lem:main}}

\noindent Fix \(r<r_e\) and set \(\tau=\lfloor nr\rfloor\). For all
\(t\le \tau<\tau_e\), the autoregressive coefficient equals one under
Model~\eqref{eq:mainmodel}. Hence,
\[
X_t
=
X_0+\sum_{j=1}^{t}u_j,
\qquad
u_j=\sigma_j\varepsilon_j,
\qquad
\varepsilon_j\stackrel{\mathrm{i.i.d.}}{\sim}\mathcal N(0,1).
\]
Define the normalized partial-sum process
\[
W_n(s)
:=
\frac{1}{\sqrt{n\,m_{n,n}}}
\sum_{t=1}^{\lfloor ns\rfloor}u_t,
\qquad s\in[0,1].
\]
Conditional on the volatility path, \(W_n\) is a centered Gaussian process with
independent increments and conditional variance function
\(
V_n(s)
:=
\langle W_n\rangle(s)
=
\frac{1}{n\,m_{n,n}}
\sum_{t=1}^{\lfloor ns\rfloor}\sigma_t^2 .
\)
By Assumption~\ref{ass:LLNscale},
\(
\sup_{s\in[0,1]}
\left|V_n(s)-s\right|
\xrightarrow{p}0.
\)
Therefore, conditional on the volatility path, \(W_n\) is asymptotically a
Brownian motion with linear time scale, and hence
\(
W_n(\cdot)\Rightarrow W(\cdot)\,\text{in }D[0,1],
\)
where \(W\) is standard Brownian motion. Since
\(
X_{\lfloor nr\rfloor}
=
X_0+\sum_{t=1}^{\lfloor nr\rfloor}u_t,
\)
dividing by \(\sqrt{n\,m_{n,n}}\) and using
\(X_0=o_p(\sqrt{n\,m_{n,n}})\) gives
\[
\frac{X_{\lfloor nr\rfloor}}{\sqrt{n\,m_{n,n}}}
=
\frac{X_0}{\sqrt{n\,m_{n,n}}}
+
W_n(r)
\Rightarrow
W(r).
\]
\subsection{Proof of Theorem~\ref{thm:orig-consistency}}

\noindent
Fix \(r\in(r_e,r_f)\) and let \(\tau=\lfloor nr\rfloor\). Throughout this
proof, all statements are uniform over compact subintervals of
\((r_e,r_f)\). Under Model~\eqref{eq:mainmodel},
\[
X_t
=
X_{t-1}\mathbf 1\{t<\tau_e\}
+
\delta_n X_{t-1}\mathbf 1\{t\ge \tau_e\}
+
u_t,
\qquad
u_t=\sigma_t\varepsilon_t,
\qquad
\delta_n=1+\frac{c}{n^\alpha}.
\]
Thus, for \(\tau_f > t\ge \tau_e\),
\[
X_t
=
\sum_{j=0}^{t-\tau_e}\delta_n^{\,j}u_{t-j}
+
\delta_n^{\,t-\tau_e+1}X_{\tau_e-1}.
\]
Equivalently,
\[
\delta_n^{-(t-\tau_e)}X_t
=
\sum_{j=0}^{t-\tau_e}\delta_n^{\,j-(t-\tau_e)}u_{t-j}
+
\delta_n X_{\tau_e-1}.
\]

\noindent
Define
\[
S_{n,t}
:=
\sum_{j=0}^{t-\tau_e}\delta_n^{\,j-(t-\tau_e)}u_{t-j}.
\]
Conditional on the volatility path,
\[
\E\!\left(S_{n,t}^2\mid\sigma\right)
=
\sum_{k=0}^{t-\tau_e}\delta_n^{-2k}\sigma_{t-k}^2.
\]
Taking expectations and using the uniform volatility-scale bound implied by
Assumption~\ref{ass:smooth},
we obtain
\[
\E\!\left(S_{n,t}^2\right)
=
\sum_{k=0}^{t-\tau_e}\delta_n^{-2k}\E(\sigma_{t-k}^2)
\lesssim
m_{n,n}\sum_{k=0}^{\infty}\delta_n^{-2k}
=
m_{n,n}\frac{1}{1-\delta_n^{-2}}
\sim
\frac{m_{n,n}}{2c}\,n^\alpha.
\]
Therefore,
\[
S_{n,t}
=
O_p\!\left(\sqrt{m_{n,n}}\,n^{\alpha/2}\right),
\]
and hence
\[
\frac{S_{n,t}}{\sqrt{n\,m_{n,n}}}
=
O_p\!\left(n^{(\alpha-1)/2}\right)
=
o_p(1),
\]
because \(\alpha\in(0,1)\). It follows that, for
\(t=\lfloor ns\rfloor\) with fixed \(r_f > s>r_e\),
\[
\frac{\delta_n^{-(t-\tau_e)}X_t}{\sqrt{n\,m_{n,n}}}
=
\frac{\delta_n X_{\tau_e-1}}{\sqrt{n\,m_{n,n}}}
+o_p(1).
\]
By Lemma~\ref{lem:main},
\[
\frac{X_{\tau_e}}{\sqrt{n\,m_{n,n}}}
\Rightarrow
B(r_e),
\]
where \(B\) is standard Brownian motion. Hence, for fixed \(r_f > s>r_e\),
\begin{equation}\label{eq:exp_regime_xt}
X_{\lfloor ns\rfloor}
=
\delta_n^{\,\lfloor ns\rfloor-\tau_e}
X_{\tau_e}\{1+o_p(1)\}.
\end{equation}

\noindent
Next,
\[
\frac{\sum_{j=1}^{\tau_e}X_j}{\sqrt{m_{n,n}}\,n^{3/2}}
\Rightarrow
\int_0^{r_e}B(u)\,du.
\]
This pre-bubble sum is negligible relative to the explosive contribution after
\(\tau_e\). Using \eqref{eq:exp_regime_xt} and the above panel,
\begin{align}\label{eq:exp_regime_sum_xt}
\frac{1}{\tau}\sum_{j=\tau_e}^{\tau}X_j
&=
\frac{1}{\tau}\sum_{k=0}^{\tau-\tau_e}
\delta_n^{\,k}X_{\tau_e}\{1+o_p(1)\} \notag\\
&=
\frac{n^\alpha\,\delta_n^{\,\tau-\tau_e}}{\tau c}\,
X_{\tau_e}\{1+o_p(1)\}.
\end{align}
Therefore,
\begin{equation}\label{eq:equation12}
\bar X_\tau
:=
\frac{1}{\tau}\sum_{j=1}^{\tau}X_{j-1}
=
\frac{n^\alpha\,\delta_n^{\,\tau-\tau_e}}{\tau c}\,
X_{\tau_e}\{1+o_p(1)\}.
\end{equation}

\noindent
Define
\(
\widetilde X_{j-1}:=X_{j-1}-\bar X_\tau.
\)
From \eqref{eq:exp_regime_xt} and \eqref{eq:equation12}, for
\(j\ge \tau_e\),
\[
\widetilde X_{j-1}
=
\left(
\delta_n^{\,j-1-\tau_e}
-
\frac{n^\alpha\,\delta_n^{\,\tau-\tau_e}}{\tau c}
\right)
X_{\tau_e}\{1+o_p(1)\},
\]
whereas for \(j<\tau_e\),
\[
\widetilde X_{j-1}
=
-
\frac{n^\alpha\,\delta_n^{\,\tau-\tau_e}}{\tau c}
X_{\tau_e}\{1+o_p(1)\}.
\]

\noindent
Let
\(
C_{n,\tau}:=
\frac{n^\alpha\,\delta_n^{\,\tau-\tau_e}}{\tau c}.
\) Then,
\[
\sum_{j=1}^{\tau}\widetilde X_{j-1}^{\,2}
=
X_{\tau_e}^2
\left[
\sum_{j=\tau_e}^{\tau}
\left(\delta_n^{\,j-1-\tau_e}-C_{n,\tau}\right)^2
+
\tau_e C_{n,\tau}^2
\right]\{1+o_p(1)\}.
\]
Now,
\[
\sum_{j=\tau_e}^{\tau}\delta_n^{\,2(j-1-\tau_e)}
\sim
\frac{n^\alpha}{2c}\,
\delta_n^{\,2(\tau-\tau_e)},
\]
while
\[
C_{n,\tau}\sum_{j=\tau_e}^{\tau}\delta_n^{\,j-1-\tau_e}
=
O\!\left(n^{2\alpha-1}\delta_n^{\,2(\tau-\tau_e)}\right)
=
o\!\left(n^\alpha\delta_n^{\,2(\tau-\tau_e)}\right),
\]
and
\[
\tau C_{n,\tau}^2
=
O\!\left(n^{2\alpha-1}\delta_n^{\,2(\tau-\tau_e)}\right)
=
o\!\left(n^\alpha\delta_n^{\,2(\tau-\tau_e)}\right),
\]
because \(\alpha<1\). Hence,
\begin{equation}\label{eq:equation15}
\sum_{j=1}^{\tau}\widetilde X_{j-1}^{\,2}
=
\frac{n^\alpha}{2c}\,
\delta_n^{\,2(\tau-\tau_e)}X_{\tau_e}^2
\{1+o_p(1)\}.
\end{equation}

\noindent
Also we note that,
\begin{align}\label{eq:signal-cross}
\sum_{j=\tau_e}^{\tau}\widetilde X_{j-1}X_{j-1}
&=
X_{\tau_e}^2
\sum_{j=\tau_e}^{\tau}
\left(\delta_n^{\,j-1-\tau_e}-C_{n,\tau}\right)
\delta_n^{\,j-1-\tau_e}
\{1+o_p(1)\} \notag\\
&=
\frac{n^\alpha}{2c}\,
\delta_n^{\,2(\tau-\tau_e)}X_{\tau_e}^2
\{1+o_p(1)\}.
\end{align}
Thus the signal part of the numerator has the same leading order as the
denominator.

\noindent
The recursive least-squares estimator satisfies
\[
\widehat\delta_\tau-1
=
\frac{\sum_{j=1}^{\tau}\widetilde X_{j-1}(X_j-X_{j-1})}
     {\sum_{j=1}^{\tau}\widetilde X_{j-1}^{\,2}}.
\]
Since
\(
X_j-X_{j-1}
=
(\delta_n-1)X_{j-1}\mathbf 1\{j\ge\tau_e\}
+
u_j,
\)
we have
\begin{align}
\sum_{j=1}^{\tau}\widetilde X_{j-1}(X_j-X_{j-1})
&=
(\delta_n-1)
\sum_{j=\tau_e}^{\tau}\widetilde X_{j-1}X_{j-1}
+
\sum_{j=1}^{\tau}\widetilde X_{j-1}u_j.
\end{align}
The stochastic term is negligible relative to the signal. Indeed, using the
preceding expansions and the volatility-scale bounds,
\[
\sum_{j=1}^{\tau}\widetilde X_{j-1}u_j
=
O_p\!\left(
\delta_n^{\,\tau-\tau_e}|X_{\tau_e}|
\sqrt{m_{n,n}}\,n^{\alpha/2}
\right)
+
O_p\!\left(
n^{\alpha-1}\delta_n^{\,\tau-\tau_e}|X_{\tau_e}|
\sqrt{n\,m_{n,n}}
\right).
\]
Since
\(
X_{\tau_e}=O_p(\sqrt{n\,m_{n,n}})
\)
and \(\delta_n^{\tau-\tau_e}\) diverges exponentially fast for fixed
\(r>r_e\), it follows that
\[
\sum_{j=1}^{\tau}\widetilde X_{j-1}u_j
=
o_p\!\left(
(\delta_n-1)
\sum_{j=1}^{\tau}\widetilde X_{j-1}^{\,2}
\right).
\]
Combining this with \eqref{eq:equation15} and \eqref{eq:signal-cross},
\[
\widehat\delta_\tau-1
=
(\delta_n-1)\{1+o_p(1)\}.
\]
Therefore,
\begin{align}
\text{SV-ADF}_r^{\delta}
&=
\tau(\widehat\delta_\tau-1) \notag\\
&=
\tau(\delta_n-1)\{1+o_p(1)\} \notag\\
&=
n^{1-\alpha}rc\{1+o_p(1)\}
\to\infty.
\label{eq:final_rate_re_delta}
\end{align}

\noindent
We next study the order of the residual variance estimator
\[
\widehat\sigma_\tau^{\,2}
=
\frac{1}{\tau}\sum_{j=1}^{\tau}
\bigl(\widetilde X_j-\widehat\delta_\tau\widetilde X_{j-1}\bigr)^2.
\]
Since \(\widehat\delta_\tau-1=(\delta_n-1)\{1+o_p(1)\}\), the dominant
contribution comes from the pre-bubble part \(j<\tau_e\). For \(j<\tau_e\),
\[
\widetilde X_{j-1}
=
-C_{n,\tau}X_{\tau_e}\{1+o_p(1)\},
\]
and hence
\[
(\widehat\delta_\tau-1)^2
\sum_{j=1}^{\tau_e}\widetilde X_{j-1}^{\,2}
=
\frac{\tau_e}{\tau^2}
\delta_n^{\,2(\tau-\tau_e)}X_{\tau_e}^2
\{1+o_p(1)\}.
\]
After division by \(\tau\),
\begin{equation}\label{eq:sigma-rate-orig}
\widehat\sigma_\tau^{\,2}
=
\frac{\tau_e}{\tau^3}
\delta_n^{\,2(\tau-\tau_e)}X_{\tau_e}^2
\{1+o_p(1)\}.
\end{equation}
The remaining terms in the residual expansion are of smaller order by the same
bounds used above for the stochastic numerator and by
\(\widehat\delta_\tau-\delta_n=o_p(\delta_n-1)\).

\noindent
Using \eqref{eq:equation15}, \eqref{eq:final_rate_re_delta}, and
\eqref{eq:sigma-rate-orig}, we obtain
\begin{align}\label{eq:final_rate_re_t}
\text{SV-ADF}_r^{t}
&=
\left(
\frac{\sum_{j=1}^{\tau}\widetilde X_{j-1}^{\,2}}
     {\widehat\sigma_\tau^{\,2}}
\right)^{1/2}
(\widehat\delta_\tau-1) \notag\\
&=
\left(
\frac{
\displaystyle
\frac{n^\alpha}{2c}
\delta_n^{\,2(\tau-\tau_e)}X_{\tau_e}^2
}{
\displaystyle
\frac{\tau_e}{\tau^3}
\delta_n^{\,2(\tau-\tau_e)}X_{\tau_e}^2
}
\right)^{1/2}
\frac{c}{n^\alpha}\{1+o_p(1)\} \notag\\
&=
n^{1-\alpha/2}
\frac{\sqrt c\,r^{3/2}}{\sqrt{2r_e}}
\{1+o_p(1)\}.
\end{align}
Thus the \(t\)-type statistic diverges at rate \(n^{1-\alpha/2}\). Let the right-tail critical boundaries satisfy
\(
cv_{\beta_n,r}^{\delta}\to\infty,\)  \(
cv_{\beta_n,r}^{t}\to\infty,
\)
and
\(
\frac{cv_{\beta_n,r}^{\delta}}{n^{1-\alpha}}\to0,
\frac{cv_{\beta_n,r}^{t}}{n^{1-\alpha/2}}\to0
\)
uniformly over compact subintervals of \((r_e,r_f)\). Then, for every
\(\varepsilon>0\),
\[
\inf_{r\in[r_e+\varepsilon,r_f-\varepsilon]}
\frac{\text{SV-ADF}_r^{\delta}}{n^{1-\alpha}}
\ge c(r_e+\varepsilon)\{1+o_p(1)\},
\]
and
\[
\inf_{r\in[r_e+\varepsilon,r_f-\varepsilon]}
\frac{\text{SV-ADF}_r^{t}}{n^{1-\alpha/2}}
\ge
\frac{\sqrt c\,(r_e+\varepsilon)^{3/2}}{\sqrt{2r_e}}\{1+o_p(1)\}.
\]
Consequently,
\[
\Pr\!\left(
\text{SV-ADF}_r^{\delta}>cv_{\beta_n,r}^{\delta}
\text{ for some }r\in[r_e+\varepsilon,r_f-\varepsilon]
\right)\to1,
\]
and
\[
\Pr\!\left(
\text{SV-ADF}_r^{t}>cv_{\beta_n,r}^{t}
\text{ for some }r\in[r_e+\varepsilon,r_f-\varepsilon]
\right)\to1.
\]
Together with the no-spurious-crossing result on \([r_0,r_e-\varepsilon]\),
this implies
\[
\widehat r_e\xrightarrow{p}r_e.
\]

\subsection{Proof of Theorem~\ref{thm:collapse-consistency}}

\noindent We replacte the proof of Theorem~\ref{thm:orig-consistency}. Recall model~\eqref{eq:mainmodel}, consider $r >r_f$, hence for \(t=1,\dots,\lfloor nr \rfloor\),
\[
X_t
=
X_{t-1}\mathbf{1}\{t<\tau_e \ \text{or}\ t>\tau_f\}
\;+\;
\delta_n X_{t-1}\mathbf{1}\{\tau_e \le t \le \tau_f\}
\;+\;
u_t,
\qquad
\tau=\lfloor nr\rfloor .
\]

\noindent For \(t>\tau_f\),
\begin{align}\label{eq:x_tafter_rf}
X_t
&= \sum_{j=1}^{\tau_e - 1 }u_j
\;+\;
\sum_{j=\tau_f+1}^{t}u_j
\;+\;
\sum_{j=0}^{\tau_f-\tau_e}\delta_n^{\,j}u_{\tau_f-j}
\;+\;
\delta_n^{\,\tau_f-\tau_e+1}X_{\tau_e-1} \nonumber\\
&=
O_p\!\bigl(\sqrt{n\,m_{n,n}}\bigr)
\;+\;
O_p\!\bigl(n^{\alpha/2}\delta_n^{\,\tau_f-\tau_e}\sqrt{m_{n,n}}\bigr)
\;+\;
\delta_n^{\,\tau_f-\tau_e+1}X_{\tau_e-1} \nonumber \\
&=
\delta_n^{\,\tau_f-\tau_e+1}X_{\tau_e-1}\bigl(1+o_p(1)\bigr).
\end{align}

\noindent  We already know,
\[
\frac{\sum_{j=1}^{\tau_e}X_j}{\sqrt{m_{n,n}}\,n^{3/2}}
\ \Rightarrow\ \int_0^{r_e} B(u)\,du.
\]
Combining the earlier panel with \eqref{eq:exp_regime_sum_xt},
\begin{align*}
\frac{1}{\tau}\sum_{j=1}^{\tau}X_j
&=
\frac{1}{\tau}\sum_{j=1}^{\tau_e}X_j
+
\frac{1}{\tau}\sum_{k=0}^{\tau_f-\tau_e}\delta_n^{\,k}X_{\tau_e}\bigl(1+o_p(1)\bigr)
+
\frac{1}{\tau}\sum_{j=\tau_f+1}^{\tau}
\delta_n^{\,\tau_f-\tau_e+1}X_{\tau_e-1}\bigl(1+o_p(1)\bigr) \\
&=
O_p\!\left(\frac{\sqrt{m_{n,n}}\,n^{3/2}}{\tau}\right)
+
\frac{n^\alpha\delta_n^{\,\tau_f-\tau_e}}{\tau c}\,
X_{\tau_e}\bigl(1+o_p(1)\bigr)
+
\frac{\tau-\tau_f}{\tau}\,
\delta_n^{\,\tau_f-\tau_e}X_{\tau_e}\bigl(1+o_p(1)\bigr) \\
&=
\frac{r-r_f}{r}\,
\delta_n^{\,\tau_f-\tau_e}X_{\tau_e}\bigl(1+o_p(1)\bigr).
\end{align*}
Define
\(
a:=\frac{r-r_f}{r}.
\)
Then, for all \(\tau_e\le t\le \tau_f\),
\[
\widetilde X_t
=
\left(
\delta_n^{\,t-\tau_e}
-a\,\delta_n^{\,\tau_f-\tau_e}
\right)X_{\tau_e}\bigl(1+o_p(1)\bigr),
\] so that,
\begin{align}\label{eq:equation26}
\sum_{t=\tau_e}^{\tau_f}\widetilde X_t^{\,2}
&=
\sum_{t=\tau_e}^{\tau_f}
\left(
\delta_n^{\,t-\tau_e}
-a\,\delta_n^{\,\tau_f-\tau_e}
\right)^2
X_{\tau_e}^2\bigl(1+o_p(1)\bigr) \nonumber \\
&=
a^2(\tau_f-\tau_e)\,
\delta_n^{\,2(\tau_f-\tau_e)}X_{\tau_e}^2\bigl(1+o_p(1)\bigr).
\end{align}
Similarly, for all \(t<\tau_e\),
\(
\widetilde X_t
=
-a\,\delta_n^{\,\tau_f-\tau_e}X_{\tau_e}\bigl(1+o_p(1)\bigr),
\)
while for all \(t>\tau_f\),
\[
\widetilde X_t
=
(1-a)\,\delta_n^{\,\tau_f-\tau_e}X_{\tau_e}\bigl(1+o_p(1)\bigr).
\]
Hence,
\begin{align}\label{eq:equation27}
\sum_{j\notin[\tau_e,\tau_f]}\widetilde X_{j-1}^{\,2}
&=
\sum_{j<\tau_e}\widetilde X_{j-1}^{\,2}
+
\sum_{j>\tau_f}\widetilde X_{j-1}^{\,2} \nonumber \\
&=
\Bigl(
a^2\tau_e+(\tau-\tau_f)(1-a)^2
\Bigr)
\delta_n^{\,2(\tau_f-\tau_e)}X_{\tau_e}^2\bigl(1+o_p(1)\bigr).
\end{align}

\noindent Thus combining \eqref{eq:equation26} and \eqref{eq:equation27},
\begin{equation}\label{eq:tilde_after_tf}
    \sum_{t=1}^{\tau}\widetilde X_t^{\,2}
=
\Bigl(a^2\tau_f+(\tau-\tau_f)(1-a)^2\Bigr)
\delta_n^{\,2(\tau_f-\tau_e)}X_{\tau_e}^2\bigl(1+o_p(1)\bigr).
\end{equation}

\noindent Next,
\[
\sum_{j=1}^{\tau}\widetilde X_{j-1}(X_j-X_{j-1})
=
\sum_{j=1}^{\tau_e}\widetilde X_{j-1}u_j
+
\sum_{j=\tau_f+1}^{\tau}\widetilde X_{j-1}u_j
+
\sum_{j=\tau_e+1}^{\tau_f}\widetilde X_{j-1}
\left(
u_j+\frac{c}{n^\alpha}X_{j-1}
\right).
\]
Using the preceding expansions,
\begin{align}\label{eq:cross_tilde_fater_tf}
\sum_{j=1}^{\tau}\widetilde X_{j-1}(X_j-X_{j-1})
&=
\sum_{j=1}^{\tau_e}\Bigl(-a\,\delta_n^{\,\tau_f-\tau_e}X_{\tau_e}\Bigr)u_j\bigl(1+o_p(1)\bigr) \nonumber \\
&\quad
+\sum_{j=\tau_f+1}^{\tau}(1-a)\,\delta_n^{\,\tau_f-\tau_e}X_{\tau_e}\,u_j\bigl(1+o_p(1)\bigr) \nonumber \\
&\quad
+\sum_{j=\tau_e+1}^{\tau_f}
\Bigl(\delta_n^{\,j-1-\tau_e}-a\,\delta_n^{\,\tau_f-\tau_e}\Bigr)
\left(
u_j+\frac{c}{n^\alpha}X_{j-1}
\right)X_{\tau_e}\bigl(1+o_p(1)\bigr) \nonumber \\
&=
O_p\!\left(\delta_n^{\,\tau_f-\tau_e}X_{\tau_e}\sqrt{n\,m_{n,n}}\right) \nonumber \\
&\quad
+\frac{c}{n^\alpha}
\sum_{j=\tau_e+1}^{\tau_f}
\Bigl(\delta_n^{\,j-1-\tau_e}-a\,\delta_n^{\,\tau_f-\tau_e}\Bigr)
X_{j-1}X_{\tau_e}\bigl(1+o_p(1)\bigr) \nonumber \\
&=
\left(\frac12-a\right)\delta_n^{\,2(\tau_f-\tau_e)}X_{\tau_e}^2\bigl(1+o_p(1)\bigr).
\end{align}
Combining \eqref{eq:tilde_after_tf} and \eqref{eq:cross_tilde_fater_tf},
\begin{equation}\label{eq:delta_rate_after_tf}
    \tau\bigl(\widehat\delta_\tau-1\bigr)
=
\tau\,
\frac{\sum_{j=1}^{\tau}\widetilde X_{j-1}(X_j-X_{j-1})}
     {\sum_{j=1}^{\tau}\widetilde X_{j-1}^{\,2}}
=
\frac{r^2\bigl(\tfrac12-a\bigr)\bigl(1+o_p(1)\bigr)}
     {r_f(r-r_f)}
=
O_p(1).
\end{equation}
 Now let \(\tau=\lfloor nr\rfloor\) with \(r\in[r_0,1]\), and similarly define,
\[
\widehat\sigma_\tau^{\,2}
=
\frac{1}{\tau}\sum_{j=1}^{\tau}
\bigl(\widetilde X_j-\widehat\delta_\tau\,\widetilde X_{j-1}\bigr)^2,
\qquad
u_j
:=
X_j-X_{j-1}
-\bigl(\delta_n-1\bigr)X_{j-1}\mathbf{1}\{\tau_e\le j\le \tau_f\}.
\]
Then,
\[
\widehat\sigma_\tau^{\,2}
=
\frac{1}{\tau}\sum_{j=1}^{\tau}
\Bigl[
u_j
-(\widehat\delta_\tau-\delta_n)\widetilde X_{j-1}\mathbf{1}\{\tau_e\le j\le \tau_f\}
-(\widehat\delta_\tau-1)\widetilde X_{j-1}\mathbf{1}\{j<\tau_e \text{ or } j>\tau_f\}
\Bigr]^2.
\]
Expanding the square gives
\begin{align}\label{eq:sigma_hat}
\widehat\sigma_\tau^{\,2}
&=
\frac{1}{\tau}\sum_{j=1}^{\tau}u_j^2
-\frac{2}{\tau}\sum_{j=\tau_e}^{\tau_f}
(\widehat\delta_\tau-\delta_n)\widetilde X_{j-1}u_j
+\frac{1}{\tau}\sum_{j=\tau_e}^{\tau_f}
(\widehat\delta_\tau-\delta_n)^2\widetilde X_{j-1}^2 \nonumber\\
&\qquad
-\frac{2}{\tau}\sum_{j\notin[\tau_e,\tau_f]}
(\widehat\delta_\tau-1)\widetilde X_{j-1}u_j
+\frac{1}{\tau}\sum_{j\notin[\tau_e,\tau_f]}
(\widehat\delta_\tau-1)^2\widetilde X_{j-1}^2 \nonumber \\
&\sim
\tau^{-1}\sum_{j=1}^{\tau}u_j^2
+
\frac{1}{\tau n^{2\alpha}}
\sum_{j=\tau_e}^{\tau_f}\widetilde X_{j-1}^{\,2}
+
\frac{1}{\tau n^\alpha}
\sum_{j=\tau_e}^{\tau_f}\widetilde X_{j-1}u_j \nonumber \\
&\qquad
+
\frac{n^{-2}}{\tau}
\sum_{j\notin[\tau_e,\tau_f]}\widetilde X_{j-1}^{\,2}
+
\frac{1}{n\tau}
\sum_{j\notin[\tau_e,\tau_f]}\widetilde X_{j-1}u_j .
\end{align}

Note that the limit on the right-hand side of \eqref{eq:delta_rate_after_tf} features a nontrivial factor determined by the true bubble window. The statistic remains \(O_p(1)\) after \(\tau_f\) because the post-collapse unit-root segment shifts the recursive sample mean by an amount of the same order as the collapsed bubble level. Thus the demeaned process no longer has the explosive accumulation that drives the statistic to infinity before \(\tau_f\). Consequently, the SV-ADF statistic stays of order \(1\) beyond \(\tau_f\).
\noindent Hence, from \eqref{eq:sigma_hat}
\begin{equation}\label{eq:rate-sigma_hat}
    \widehat\sigma_\tau^{\,2}
\sim
n^{-2\alpha}\delta_n^{\,2(\tau_f-\tau_e)}X_{\tau_e}^2\bigl(1+o_p(1)\bigr).
\end{equation}

\noindent Therefore, combining \eqref{eq:tilde_after_tf} and \eqref{eq:rate-sigma_hat},
\begin{equation}\label{eq:equation33}
    \text{SV-ADF}_r^t = \left(
\frac{\sum_{j=1}^{\tau}\widetilde X_{j-1}^{\,2}}
     {\widehat\sigma_\tau^{\,2}}
\right)^{1/2}
(\widehat\delta_\tau-1)
\sim
n^{\alpha-1/2}.
\end{equation}

\noindent The limit theory for the terminal estimate \(\widehat r_f\) under the alternative now follows. First, for all \(\tau=\lfloor nr\rfloor\) with \(r>r_f\),
\[
\Pr\!\left(
\text{SV-ADF}_r^{\delta}
<
cv_{\beta_n,l}^{\delta}
\right)
=
\Pr\!\left(
\frac{r^2\bigl(\tfrac12-a\bigr)\bigl(1+o_p(1)\bigr)}
     {r_f(r-r_f)}
<
cv_{\beta_n,l}^{\delta}
\right)
\to 1,
\]
provided
\(
cv_{\beta_n,l}^{\delta}\to \infty.
\)
Moreover, to obtain consistency of \(\widehat r_f\), one must also rule out premature downward crossings on the still-explosive interval \(r_e<r<r_f\). On that interval,
\[
\text{SV-ADF}_r^{\delta}
=
n^{1-\alpha}cr\{1+o_p(1)\},
\]
so it is sufficient that
\(
\frac{cv_{\beta_n,l}^{\delta}}{n^{1-\alpha}}\to 0.
\)
Hence, the coefficient-based collapse date is consistently estimated provided
\(
\frac{1}{cv_{\beta_n,l}^{\delta}}
\;+\;
\frac{cv_{\beta_n,l}^{\delta}}{n^{1-\alpha}}
\to 0.
\) Similarly for $r > r_f$,
\[
\Pr\!\left(
\text{SV-ADF}_r^{t}
<
cv_{\beta_n,l}^{\mathrm{t}}
\right)
=
\Pr\!\left(
n^{\alpha-1/2}
<
cv_{\beta_n,l}^{\mathrm{t}}
\right)
\to 1,
\]
which holds provided
\(
\frac{n^{\alpha-1/2}}{cv_{\beta_n,l}^{\mathrm{t}}}\to 0.
\)
To rule out downward crossings for \(r_e<r<r_f\), note that on this interval the \(t\)-type statistic is of order \(n^{1-\alpha/2}\), so it is sufficient that
\(
\frac{cv_{\beta_n,l}^{\mathrm{t}}}{n^{1-\alpha/2}}\to 0.
\)
Therefore, consistency of the \(t\)-type collapse date follows, provided that
\[
\frac{n^{\alpha-1/2}}{cv_{\beta_n,l}^{\mathrm{t}}}
\;+\;
\frac{cv_{\beta_n,l}^{\mathrm{t}}}{n^{1-\alpha/2}}
\to 0.
\]
Hence, the proof is concluded from similar uniformity argument as in the bubble origination case.

\subsection{Plots for $X_{\tau_f} -X_{\tau_e} \neq O_p(1)$}\label{sec:pwy_contra}

\begin{figure}[!htpb]
\centering
\includegraphics[width=\linewidth]{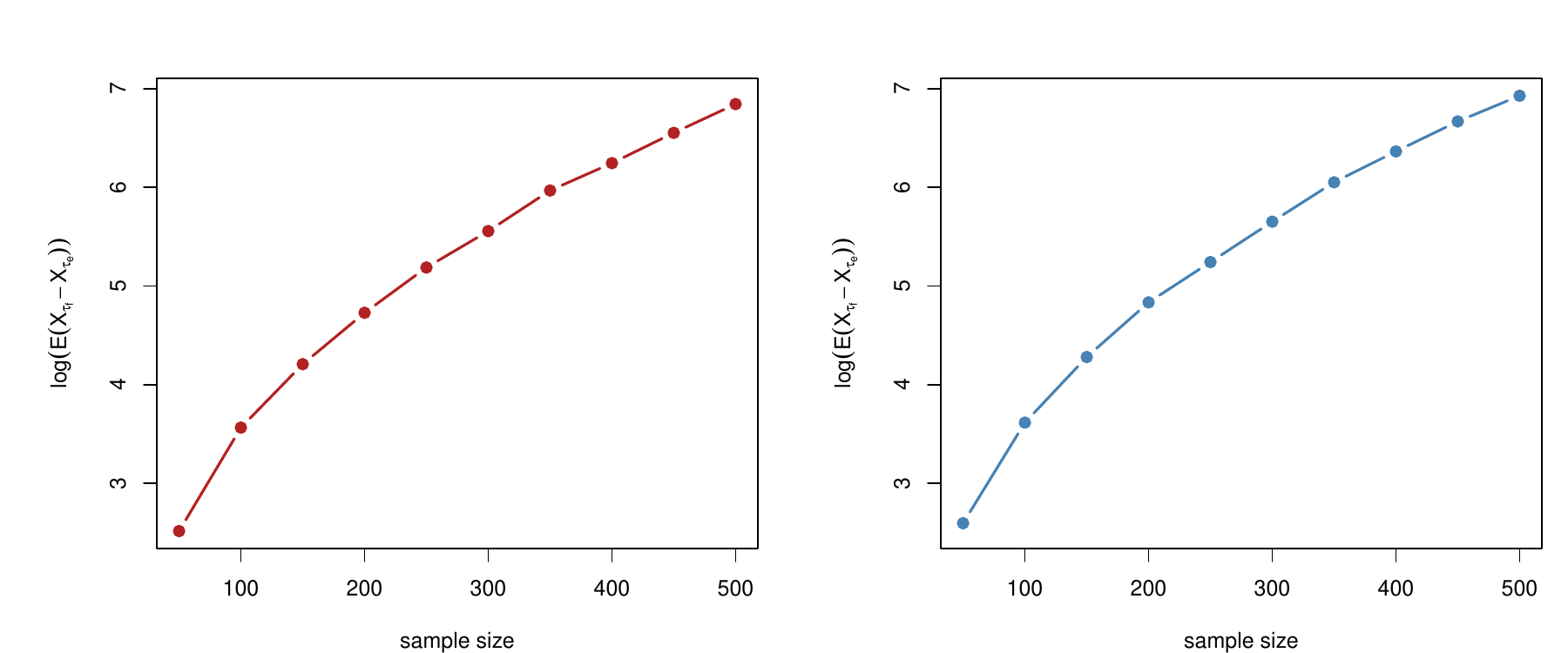}   
\captionsetup{skip=0pt}
\caption{\footnotesize{The panels report the average value of \(X_{\tau_f}-X_{\tau_e}\) on a logarithmic scale under homoskedasticity (left) and stochastic volatility (right). In both panels, \(r_e=0.4\), \(r_f=0.6\), \(c=1\), and \(\alpha=0.5\). The sample size ranges from \(n=50,100,\ldots,500\), while the number of Monte Carlo replications is fixed at \(B=500\). The visual evidence indicates that this difference grows exponentially with the sample size, which stands in sharp contradiction to the PWY maintained assumption.}  }
\label{fig:pwy_contradiction}
\end{figure}

Using simulation evidence, Figure~\ref{fig:pwy_contradiction} complements the theoretical point that a reset of order  \(O_p(1)\) is implausible to reconcile with the magnitudes generated in finite samples under a mildly explosive episode.

\subsection{Bubble detection in additional Crypto-currencies}\label{sec:more_crypto}

\begin{figure}[H]
\centering
\includegraphics[width=\linewidth]{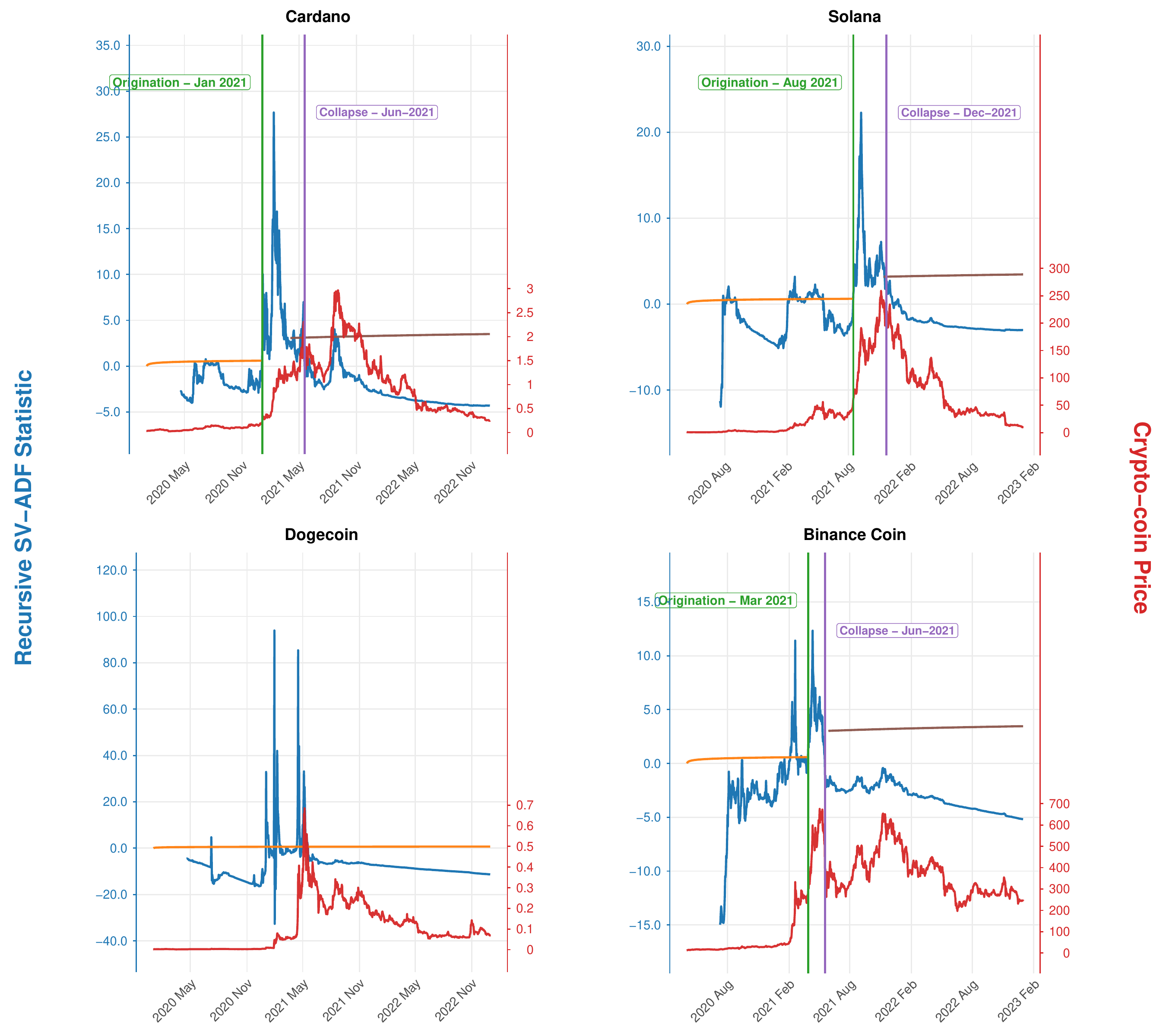}   
\captionsetup{skip=0pt}
\caption{\footnotesize  The recursive SV-ADF test indicates a pronounced bubble episode in Cardano and Solana during 2021. By contrast, Binance Coin exhibited only weak and short-lived exuberance, while Dogecoin showed no evidence of a bubble.}
\label{fig:crypto_4}
\end{figure}

In Figure~\ref{fig:crypto}, we applied our date-stamping procedure to Bitcoin and Ethereum and found that the standard PWY procedure under homoskedasticity fails to detect accurate date-stamps in those cryptocurrencies. Accounting for volatility dynamics also reveals statistically significant bubble behavior in several other major cryptoassets, as illustrated in Figure~\ref{fig:crypto_4}.

\subsection{Threshold selections: $cv_{\beta,r}$ and $ cv_{\beta,l}$}\label{sec:threshs}

\begin{table}[H]
\centering
\setlength{\tabcolsep}{6pt}
\begin{tabular}{c c c}
\toprule
\(n\) & \(cv^{\delta}_{0.10,H_0}\) & \(cv^{\delta}_{0.10,H_1}\) \\
\midrule
500  & 0.6463 & -0.2566 \\
550  & 0.5563 & -0.9916 \\
600  & 0.6304 & 1.2746 \\
650  & 0.7157 & 1.9418 \\
700  & 0.7097 & 1.0183 \\
750  & 0.7037 & 2.6310 \\
800  & 0.7823 & 3.2163 \\
850  & 0.7888 & 3.8318 \\
900  & 0.8174 & 4.6070 \\
950  & 0.8374 & 3.2394 \\
1000 & 0.8403 & 4.1479 \\
\bottomrule
\end{tabular}
\captionsetup{skip=1pt}
\caption{Simulated coefficient-based critical values under \(H_0\) and \(H_1\) for \(n=500,550,\ldots,1000\), a range relevant for our two- to four-year empirical windows, corresponding to about 500--1000 trading days. Based on \(B=1000\) replications, the 90\% upper quantile under \(H_0\) is well approximated by \(\log(nr)/10\), while the 10\% lower quantile under \(H_1\) is in the range of \(\log(nr)/2\). We adopt these cutoffs in alignment with the asymptotic theory developed in the main paper.}
\label{tab:cv_coeff_h0_h1_500_1000}
\end{table}

\begin{figure}[!htpb]
\centering
\includegraphics[width=\linewidth]{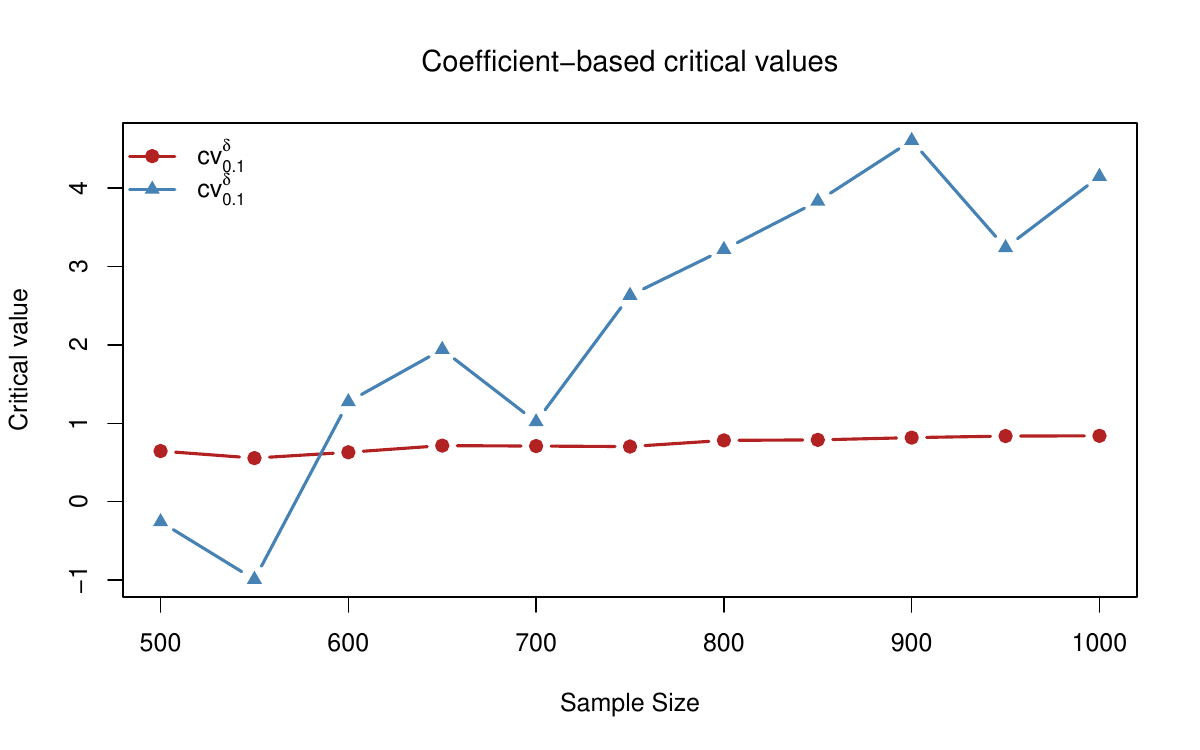}   
\captionsetup{skip=0pt}
\caption{\footnotesize{ Threshold plots under $H_0$ (red) and $H_1$ (blue) in the settings as discussed in Table~\ref{tab:cv_coeff_h0_h1_500_1000}}.}
\label{fig:critical_values}
\end{figure}
\subsection{GARCH: Double Local-to-Unity Behavior }\label{sec:garch}
As an alternative to the log-AR volatility specification, one may consider a zero-intercept near-integrated GARCH(1,1) process under the same mildly explosive mean dynamics:
\[
X_{t,n}=\delta_n X_{t-1,n}+u_{t,n},
\qquad
\delta_n=1+\frac{c}{n^\alpha},
\qquad c>0,\ \alpha\in(0,1),
\]
with
\[
u_{t,n}=\sigma_{t,n}\varepsilon_t,
\qquad
\varepsilon_t\stackrel{iid}{\sim}\mathcal N(0,1),
\]
and
\[
\sigma_{t,n}^2=\alpha_n u_{t-1,n}^2+\beta_n\sigma_{t-1,n}^2,
\qquad
\alpha_n,\beta_n\ge 0,
\qquad
\sigma_{0,n}^2>0.
\]
To capture double local-to-unity behavior, we assume
\[
\alpha_n+\beta_n=1-\kappa_n<1,
\qquad
\kappa_n\downarrow 0,
\qquad
n\kappa_n\to 0,
\]
so that volatility persistence approaches unity from below sufficiently fast to keep
\[
\E(\sigma_{t,n}^2)=(\alpha_n+\beta_n)^t\sigma_{0,n}^2
\]
asymptotically nondegenerate over \(t\le n\). Since higher-order arguments require moment control, we additionally impose
\[
3\alpha_n^2+2\alpha_n\beta_n+\beta_n^2<1,
\]
which, under \(\alpha_n+\beta_n=1-\kappa_n\), is equivalent to $
(1-\kappa_n)^2+2\alpha_n^2<1,$
and hence may be enforced by the sufficient rate condition:
\[
\alpha_n^2=o(\kappa_n).
\]

\newpage
\bibliographystyle{apalike}

\newpage

\end{spacing}    
\end{document}